\documentclass[10pt, twocolumn,article]{IEEEtran}

\usepackage[utf8]{inputenc}
\usepackage{cite}
\usepackage{setspace}
\usepackage{epsf}\epsfverbosetrue
\usepackage{graphics,epsfig}
\usepackage{graphicx,epsfig}
\usepackage{epsfig}
\usepackage{multirow}
\usepackage{alltt}
\usepackage{subfigure}
\usepackage{tcolorbox}
\usepackage{float}
\usepackage{url}
\usepackage{graphicx}
\usepackage{verbatim} 
\usepackage{footnote} 
\usepackage{latexsym} 
\usepackage{sidecap} 
\usepackage{wrapfig}
\usepackage{eso-pic}
\usepackage{fix-cm}
\usepackage{algorithm}
\usepackage{algorithmic}
\usepackage{color}
\usepackage{wrapfig}
\usepackage{multicol}
\usepackage{soul}

\usepackage{pifont}
\usepackage{flowchart}
\usetikzlibrary{shapes,arrows,matrix,decorations.pathreplacing,shapes.geometric,positioning,calc}  

\usepackage{dsfont}
\usepackage{amssymb}
\usepackage{array}
\usepackage{xcolor}

\usepackage{layout}

\usepackage{amsmath}
\usepackage[
  locale = DE 
]{siunitx}

\usepackage{textcomp}
\usepackage{multirow}
\usepackage{mathtools}
\usepackage{soul} 
\usepackage{amsmath} 
\usepackage{verbatim}
\usepackage{csvsimple} 
\usepackage{array}
\usepackage{subfig}
\usepackage[normalem]{ulem}
\usepackage{booktabs}
\newcolumntype{P}[1]{>{\RaggedRight\arraybackslash}p{#1}}
\newcommand{\tabitem}{\textbullet~~}

\usepackage{makecell}
\setcellgapes{2pt}

\newcommand{\tns}[1]{\textcolor{black}{#1}}
\newcommand{\jpdc}[1]{\textcolor{black}{#1}}

\newcommand{\cont}[1]{\textcolor{black}{#1}}

\begin{document}

\title{Differential Privacy in Blockchain Technology:\\ A Futuristic Approach}

\author{Muneeb~Ul~Hassan,~Mubashir~Husain~Rehmani,~and~Jinjun~Chen%
\IEEEcompsocitemizethanks{\IEEEcompsocthanksitem M. Ul Hassan and J. Chen are with the Swinburne University of Technology, Hawthorn VIC 3122, Australia (e-mail: muneebmh1@gmail.com;
jinjun.chen@gmail.com)\protect\\
\IEEEcompsocthanksitem M.H. Rehmani is with the Department of Computer Science, Cork Institute of Technology, Rossa Avenue, Bishopstown, Cork, Ireland (email: mshrehmani@gmail.com).}}

\maketitle

\begin{abstract}

Blockchain has received a widespread attention because of its decentralized, tamper-proof, and transparent nature. Blockchain works over the principle of distributed, secured, and shared ledger, which is used to record, and track data within a decentralized network. This technology has successfully replaced certain systems of economic transactions in organizations and has the potential to overtake various industrial business models in future. Blockchain works over peer-to-peer (P2P) phenomenon for its operation and does not require any trusted-third party authorization for data tracking and storage. The information stored in blockchain is distributed throughout the decentralized network and is usually protected using cryptographic hash functions. Since the beginning of blockchain technology, its use in different applications is increasing exponentially, but this increased use has also raised some questions regarding privacy and security of data being stored in it. Protecting privacy of blockchain data using data perturbation strategy such as differential privacy could be a novel approach to overcome privacy issues in blockchain. In this article, we cover the topic of  integration of differential privacy in each layer of blockchain and in certain blockchain based scenarios. Moreover, we highlight some future challenges and application scenarios in which integration of differential privacy in blockchain can produce fruitful results. 

\end{abstract}



\tikzstyle{decision} = [diamond, draw, fill=blue!50]
\tikzstyle{line} = [draw, -stealth, line width= 0.4mm]
\tikzstyle{elli}=[draw, ellipse, fill=red!50,minimum height=5mm, text width=5em, text centered]
\tikzstyle{block} = [draw, rectangle, fill=red!45, rounded corners, minimum height= 10mm, minimum width = 20mm, text width=11em, text centered]

\tikzstyle{medblock} = [draw, rectangle, fill=blue!40, rounded corners, minimum height= 10mm, text width=7em, text centered]



\section{Blockchain and Differential Privacy: A Revolutionizing Integration}

\subsection{Blockchain Technology}

Blockchain technology first emerged as a tool to manage cryptocurrency in 2008 when S. Nakamoto introduced “Bitcoin” as first P2P digital cash system using blockchain~\cite{blchref01}. Blockchain works over the phenomenon of decentralizing the \tns{system using a shared distributed ledger, which is basically a data structure which contains transactions} list in an ordered form. For instance, the ledger can record all exchanged goods in a market or can store information of transactions carried out between multiple bank accounts. After every transaction in blockchain, the information stored on distributed ledger is replicated over all the blockchain nodes~\cite{blchref02}. The ledger is capable to store large amount of data as it usually records the entire history of transactions or changes that take place among all blockchain nodes in order to backtrack any transaction. Contrary to traditional databases that have central trusted environment or trusted third-party environment to store records, blockchain works over distributed set of nodes/users. Each individual node has information regarding the transactions being carried out in the network. Similarly, each node is linked with its predecessor node by using a cryptographic pointer, that makes the transactions and sharing of information more secure.\\
These decentralized system and shared ledger functionalities of blockchain also make it an optimal choice among researchers for quick, easy, more secure, and efficient way of data exchange and storage in different ways of life. Researchers are integrating blockchain technology in certain domains of everyday life. For examples, researches are being \tns{carried out to implement blockchain in real-estate, asset management, Internet of Things (IoT), healthcare, and assisting wedding scenarios.} All of these applications show that blockchain is going to take over certain major daily life domains in future.\\
Furthermore, by considering a higher level of divisions, blockchain can be categorized into two sub domains named as public and private blockchain. However, these decentralized transactions come up with certain \tns{privacy risks and attacks that require solution before integration of blockchain in our everyday life.} In this section, we present details about operation phases, types and the privacy issues in blockchain. 
\subsubsection{Operation Phases of Blockchain}
There are certain basic phenomenon that constitute the backbone of blockchain known as operation phases. In operation phases, we discuss the consensus and mining phenomenon of blockchain.
\paragraph{Consensus and Mining in Blockchain}
In order to eliminate a trusted third party or centralized entity, a specific consensus is being followed by all nodes of blockchain so that no conflict arises in future. All nodes take part in in the consensus process and allow the transactions to be carried out in the network, the update is then replicated in the ledger and broadcast throughout the blockchain. Similarly, mining is a process to collect the transaction data, and create block to attach it to blockchain database. These blocks are also validated by all other nodes to maintain transparency in blockchain. The nodes doing mining are known as \textit{miners}, and they use their \tns{computational power to create a block as early as possible, in order to get the mining reward.} The mining reward is calculated on the basis of consensus approach used to mine the block~\cite{blchref04}. \cont{For example, there are plenty of consensus algorithms, such as proof of work (PoW), proof of importance (PoI), practical Byzantine fault tolerance (PBFT),} measure of trust (MoT), proof of stake (PoS), and proof of space (PoSpace). The technique which is used in Bitcoin and many other technologies of blockchain is PoW, in which a hard-mathematical puzzle is solved by miners to validate the transaction and win the reward. This reward is further added to the network of blockchain. Another cryptocurrency named as Ether~\cite{blchref05} follows PoW or PoS, in which pseudo-random way is used to choose the miner. \tns{Mining chance of a node depends upon the wealth/stake invested by that specific node in the network.} For instance, the more wealth a node has invested, the greater will be its chance to mine the block and get the reward. Similarly, other mentioned mechanisms for consensus are also used in few \tns{applications of blockchain in order to enhance trust in the network~\cite{blchref04}.}

\subsubsection{Privacy of Blockchain}
Blockchain technology is well-known for its secure transaction mechanism, authentication and encryption are used in blockchain are the two most important services offered by blockchain to ensure data security. These services are implied in blockchain via cryptography using public key encryption, in which the participants are required to have public and private information of keys by which they can manage respective transactions. Public key cryptography works over the principle of two types of keys; \tns{ \textit{public keys} (distributed network keys)and \textit{private keys} (secret individual keys).}\\ Blockchain-based distributed \tns{public key infrastructure (PKI) is the most common technique which provides functionality of key management} for cryptography in blockchains~\cite{chenref01}. Blockchain-based PKI approaches do not have any central point of access or trusted third-party. Furthermore, in order to ensure transparency in the public system, these approaches do not require any prior trustworthiness from the nodes or participants of the system. \cont{Several blockchain approaches for PKI encryption such as Instant Karma PKI~\cite{review01}, Pemcor~\cite{review02}, Gan’s approach~\cite{review03}, Blockstack~\cite{review04}, and Certcoin~\cite{review05} have been discussed in literature to provide secure transaction among blockchain nodes.} \tns{After analysing above discussion, it is clear} that a considerable amount of work has already been done over securing blockchain. However, the privacy aspect of blockchain is not completely addressed till now. Adopting the measures to secure blockchain are certainly valuable, but one cannot neglect the need of privacy in blockchain.\\
S. Nakamoto in~\cite{blchref01} discusses that if the identity of owner of a private key gets revealed, it can then lead to disclosure of other transactions by same owners by using linking phenomenon. Similarly, the anonymity property, which is considered to be the most important feature of blockchain can also be compromised by using certain attacks~\cite{blchref07}. Therefore, privacy preservation in applications of blockchain is an important issue that need to be addressed. Few researchers worked over enhancing the privacy of using different strategies. For example, Axon in~\cite{blchref08} discusses two-level anonymity to overcome privacy issues of blockchain.\\ Similarly, the authors in~\cite{blchref09} discussed overcoming transactional privacy (e.g., confidentiality) issues in a public blockchain to mitigate the privacy challenges and enhance trust parameter in blockchain. However, we believe that using differential privacy preservation strategy in blockchain, \tns{that uses data perturbation mechanism to protect private data can be state-of-the-art solution to resolve certain privacy issues of blockchain.}

\subsection{Differential Privacy}

The idea of preserving privacy by adding adequate amount of noise in the data was first brought into attention by C. Dwork in 2006~\cite{dpref01}. The initial notion of differential privacy was presented to protect the privacy of statistical databases by adding noise to the data before query evaluation. However, researchers started using the concepts of differential privacy in other critical domains too and found out that differential privacy serves as one of the most useful privacy preserving strategy to protect personal data. \tns{Differential privacy preserves sensitive data by adding a specific value of noise (after calculation)} to preserve individual privacy. Differential privacy guarantees that \tns{presence or absence of any specific participant in a dataset} does not affect the query output results of that database. This concept of differential privacy is further applied by researchers in various applications as well, for example real-time health data monitoring, IoT data, energy systems, etc. To further understand differential privacy, we discuss two examples, and important parameters of differential privacy in this section.


\begin{table*}[t!]
\begin{center}
 \centering
 \makegapedcells
 \footnotesize
 \captionsetup{labelsep=space}
 \captionsetup{justification=centering}
\caption{\textsc{\\ \cont{Basic Medical Record for Experimental Evaluation of Utility-Privacy Tradeoff}}}
  \label{tab:tradetable}
 {\color{black} \begin{tabular}{|c|c|c|c|c|c|c|}
  	\hline
\rule{0pt}{3ex}
\bfseries Disease & Cough & Asthma & Diabetes  & Dehydration & Headache & Malaria  \\
\hline

\rule{0pt}{3ex}
\bfseries No. of Patients & 26 & 15 & 22 & 08 & 28 & 05 \\
\hline

 \end{tabular}}
  \end{center}
\end{table*}


\subsubsection{Example}
A smart electric meter reports the data usage of Hazel’s house after every 5 minutes, the information reported to electric utility contains the units of electricity consumed in previous 5 minutes. This real-time reporting of units can provide certain useful benefits, such as real-time price querying, future demand response calculation, reducing electricity short-fall during peak hours, etc. Thus, the smart meter user Hazel can enjoy advanced power management by reporting his instantaneous usage of electricity. However, if this private data gets leaked or any adversary gets access to this data, then it can easily infer the lifestyle and daily routine of Hazel~\cite{dpref02}. Furthermore, this adversary \tns{can even extract the time of use of any particular appliances in a specific time slot by using non-intrusive load monitoring (NILM)} techniques. This privacy breach of smart meter of Hazel can easily be protected via dynamic differential privacy concepts. In this specific case scenario, integration of differential privacy with smart meter will perturb the data by adding adequate amount of noise to prevent exact real-time reporting of data, so that even if any adversary gets access to this critical data, it will not be able to infer the exact usage at a specific interval of time. Because the transmitted data is efficiently perturbed, so that it can be used for demand response, and other required tasks but the daily routine cannot be inferred from the data.

\begin{figure}[t]        
\centering
\includegraphics[scale = 0.5]{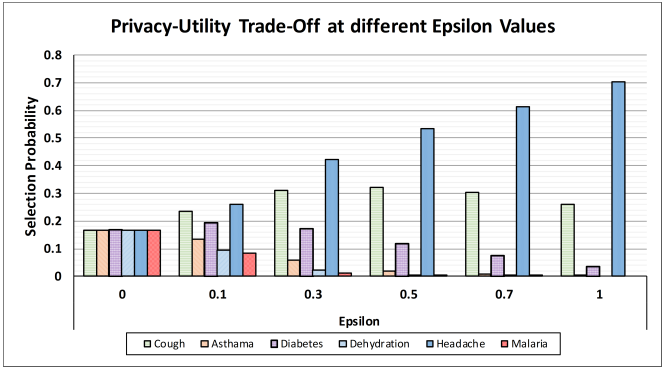}
 \caption{\cont{Trade-Off between Utility and Privacy during Query Evaluation via Differential Privacy Protection.}}      
  \label{fig:figTrade}   
\end{figure}


\subsubsection{Sensitivity and Noise Addition Mechanism}

Two important parameters to consider while including differential privacy in any data are sensitivity calibration and adequate noise calculation.\\ \textit{(i) Sensitivity} is said to be the parameter controlling the level of indistinguishability of data. For example, in a statistical database, the random value of a parameter $y$ may be 1, or may even be 10,000, thus, the domain of this parameter $y$ will be $y \in [1, 10^5]$. \cont{The determination of sensitivity value also plays an important role in determining the perturbation value. Sensitivity is data dependant parameter and it can be calculated on the basis of maximum possible difference between two values from a neighbouring datasets that differs with only one value~\cite{jpdcref18}. Formally, the definition of sensitivity function `$f$’ for two adjacent datasets ($D_1$ \& $D_2$) is as follows:}

\begin{equation}\color{black}
\Delta f_{sen} = \max_{D_1, D_2} ||f (D_1) - f (D_2)|| 
\label{eqn:eqn212}
\end{equation}
For example, an adversary sends a query to calculate aggregated value of each $y$ such as $SUM(y)$, this query might also have the particular value of the participant whose privacy is intended to be protected. In this scenario, the differential privacy mechanism calculates noise according to the standard deviation and sensitivity within the dynamic range of $10^5$. Thus, the added noise potentially hides the critical value and adversary is not able to approximate the presence or absence of a particular individual. But if the query is quite certain and to the point, then high level of noise needs to be added that requires a high level of sensitivity. However, using high sensitivity value will reduce the usefulness of data. Thus, an adequate trade-off between the privacy and truthfulness needs maintenance by adjusting value of sensitivity accordingly. Usually the value of sensitivity varies from scenario to scenario, such as applications requiring high level of privacy use large sensitivity values and vice versa. Researchers also proposed various solutions, such as choosing dynamic sensitivity values, in which the value of sensitivity will automatically vary according to the nature and requirement of analysts and data provider~\cite{dpref03}.\\
\textit{(ii) Noise Addition} mechanism is basically the protective phenomenon \tns{that calculates minimum noise value which is required to protect privacy of data.} The output magnitude of noise depends directly upon the sensitivity value. This mechanism has a base function that requires input of certain parameters to calculate noise amount. Generally, three noise addition mechanism such as Laplace mechanism, Exponential mechanism, and Gaussian mechanism are used by researchers to calculate noise value. Similar to sensitivity, exact choice of noise addition mechanism also depends upon the nature of application. For example, in case of numerical output, Laplace and Gaussian mechanisms are generally used, while Exponential mechanism is used in case of non-numerical output~\cite{dpref04}.\\
\cont{Generally differential privacy can be defined as a randomized function $F$ which satisfies probability ($P_R$) condition of  $\varepsilon$-differential privacy if any only if for two adjacent databases $D$ and $D\prime$ and for any possible output value ($O \in Range(F)$), we get the following:}
\begin{equation}\color{black}
P_R[F(D) \in O] \leq exp(\varepsilon) \times P_R[F(D\prime) \in O]
\end{equation}
\cont{In the given equation, $Range(F)$ is the maximum range value for the function $F$. Similarly, the variable $varepsilon$ is the privacy parameter which is used to control the privacy level.}

\subsubsection{\cont{Central and Local Differential Privacy}}

\cont{The formal definition of differential privacy proposed by C.Dwork focused over protection of database privacy in a centralized manner. This trend continued for a long time and majority of existing differential privacy solutions covered the aspect of centralized datasets in which centralized server is considered as a trusted entity~\cite{cutref03}. This trend successfully protects privacy from external query evaluation; however, the privacy of users is still at risk in case of the centralized server gets hacked or it start behaving maliciously~\cite{cutref04}. This discussion lead to the foundation of local differential privacy, in which noise is added locally at user to protect privacy~\cite{cutref05}. Similarly, in certain decentralized blockchain scenarios, using local differential privacy can be more viable solution rather than using centralized differential privacy because of distributed nature of blockchain network. However, in literature both works; local differential privacy based blockchain and centralized differential privacy based blockchain have been carried by researchers which basically depends upon the need of application. A detailed discussion about integration of differential privacy in blockchain based scenario has been presented in Section~\ref{newsection}.}


\begin{table*}[t!]
\begin{center}
 \centering
 \makegapedcells
 \footnotesize
 \captionsetup{labelsep=space}
 \captionsetup{justification=centering}
\caption{\textsc{\\ \cont{Summary of Previous Survey Articles Discussing Privacy Preservation in Blockchain Technology With Respect to their Target Scenario, Contribution, Important Factors, and Discussion of Differential Privacy}}}
  \label{tab:comptable}
 {\color{black} \begin{tabular}{|p{2cm}|p{0.7cm}|c|p{5cm}|p{4cm}|p{1cm}|}
  	\hline
\rule{0pt}{3ex}
\centering \bfseries Target Scenario & \centering \bfseries Ref. \# & \centering \bfseries Year & \centering \bfseries Contribution of Article & \centering \bfseries Important Factors & \bfseries Discussed \newline DP \\
\hline

\multirow{3}{*}{}

\rule{0pt}{3ex}
\centering \bfseries Bitcoin and Underlying Technologies & ~\cite{survey01} &  2018 &  Presented a comprehensive survey on security and privacy issues of Bitcoin from various perspectives, such as functionality, vulnerabilities, standard practices, etc. &  \tabitem Underlying Technologies of Bitcoin \newline
\tabitem Robustness and Feasibility of modern solutions \newline
\tabitem Privacy threats to Bitcoin users &  No \\ 
\cline{1-6}

\rule{0pt}{3ex}
\centering \bfseries Digital Cash System and Cryptocurrencies & ~\cite{survey02} & 2018 & A detailed investigation is carried out on various technologies integrated with crypto-currencies and digital cash systems. & \tabitem Anonymity and Privacy analysis \newline \tabitem Bitcoin alternatives and extensions \newline \tabitem Key Points for designing private systems & No \\ 
\cline{1-6}

\rule{0pt}{3ex}
\centering \bfseries Privacy Preservation & ~\cite{survey03} & 2019 & Carried out an extensive literature review of privacy-preserving techniques and categorised them  on basis of four major data types. & \tabitem Technical parameters and challenges for privacy preserving strategies \newline \tabitem Blockchain deployment scenarios & Partially \\ 
\cline{1-6}

\rule{0pt}{3ex}
\centering \bfseries Security Services via Blockchain & ~\cite{blchref04} &  2019 & A survey on various blockchain platforms being used as security services. & \tabitem Security services such as confidentiality, authentication, and privacy \newline \tabitem Integration of blockchain in these security services \newline \tabitem Implementation comparison & Negligible \\ 
\cline{1-6}

\rule{0pt}{3ex}
\centering \bfseries Blockchain Applications & ~\cite{newref02} &  2019 & Carried out a detailed literature review on blockchain and its applications in various domains of Internet of Things. & \tabitem Efforts being carried out to integrate blockchain in IoT. \newline \tabitem  Challenges being faced by current centralized IoT models to shift to decentralized blockchain. & No \\
\cline{1-6}

\rule{0pt}{3ex}
\centering \bfseries Privacy Protection & ~\cite{survey06} &  2019 & Analysed privacy issues of blockchain based systems along with their solutions in a detailed manner. & \tabitem Defence mechanisms \newline \tabitem Implementation of privacy preservation approaches & No \\
\cline{1-6}

\rule{0pt}{3ex}
\centering \bfseries Privacy Protection in Blockchain based IoT &  ~\cite{newref03} & 2019 & Carried out an extensive survey on privacy preservation approaches in blockchain based IoT systems from challenges and integration perspective. & \tabitem Need of privacy in blockchain \newline \tabitem Various privacy preservation approaches \newline \tabitem Future Aspects and Challenges of those approaches & Partially \\
\cline{1-6}

\rule{0pt}{3ex}
\centering \bfseries Privacy Preservation &  ~\cite{survey08} & 2019 & Presented a literature review for typical privacy preserving approaches being used in blockchain. & \tabitem Approaches being used in Cryptocurrencies & No \\
\cline{1-6}

\rule{0pt}{3ex}
\centering \bfseries Attack Surface of Blockchain &  ~\cite{survey09} & 2020 & Systematically provided a detailed survey on attack surfaces of blockchain technology by focusing mainly on public blockchain. & \tabitem Blockchain cryptography \newline \tabitem Distributed Aspect \newline \tabitem Application oriented attacks & No \\
\cline{1-6}

\rule{0pt}{3ex}
\centering \bfseries Blockchain Solutions for IoT and IIoT & ~\cite{survey10} &  2020 & A detailed analysis on security challenges of IoT and IIoT have been presented along with their blockchain based solutions. & \tabitem Cyberattacks for IIoT \newline \tabitem Tangle (IoT specific framework) \newline \tabitem Highlights research areas and directions for IIoT based blockchain & Negligible \\
\cline{1-6}

\rule{0pt}{3ex}
\centering \bfseries Differential Privacy in Blockchain & This Work &  2020 & A thorough survey on integration of differential privacy in blockchain along with extensive future challenges and directions have been presented. & \tabitem Layer based integration for differential privacy in blockchain \newline \tabitem Blockchain scenarios integrating differential privacy \newline \tabitem Future Challenges and solutions & Yes \\
\cline{1-6}

 \end{tabular}}
  \end{center}
\end{table*}


\subsection{\cont{Trade-Off of Privacy and Utility}}
\cont{Differential privacy uses the phenomenon of adding independent identically distributed (IID) noise, whether it could be perturbing the value by adding a noisy reading, or it could be randomizing the query output via exponential (or similar) mechanism~\cite{trade01}. Differential privacy protection mechanism has the ability to protect greater chunks of data from datasets and on the other hand it can also protect the data from real-time resources by adding minute value of noise via point-wise privacy~\cite{trade02}. Perturbing query value has a direct effect on a parameter called  utility (usefulness of output value), because adding large noise can reduce utility to a minimum level. This noise addition is controlled by a parameter called as epsilon $\varepsilon$, which is also known as privacy parameter. $\varepsilon$ determines the exact amount of noise needs to be added in the query output. Another significant parameter that plays its role in determining and balancing privacy-utility trade-off is delta $\delta$, which is dependent upon the input data and the difference between two individual values within dataset~\cite{jpdcref18}. Delta plays an important role in determining the perturbation value, for instance, this value determines the amount of noise with respect to difference between individual values among two neighbouring datasets~\cite{cutref06}.} \\
\cont{In order to understand the dependence of privacy-utility trade-off relation, we carried out experiment on a self-developed medical dataset containing patients of various diseases. We carried out query evaluation of the most abundant disease at various epsilon values in order to understand the fluctuations in utility. The medical data which is evaluated is given in Table~\ref{tab:tradetable} and the experimental results are given in Figure.~\ref{fig:figTrade}. From the table, if you analyse that the most abundant disease is headache, and the lest abundant is malaria. However, if one analyses the query output result at $\varepsilon$ = 0, it can clearly be seen that the chances of picking malaria and headache are the same, which means that if the value of $\varepsilon$ = 0, then privacy is maximum but utility is minimum. Moving further to $\varepsilon$ = 0.1, one can see that chances for selection of headache increase to around 25\%, and the chances of malaria being picked are reduced to approximately 8\%. Similar observation can be seen for other diseases present in the graph in Figure~\ref{fig:figTrade} and Table~\ref{tab:tradetable}. Moving further to greater $\varepsilon$ values, the chances of abundant disease getting picked is increasing and the less abundant are reducing accordingly. For example, at $\varepsilon$ = 0.5, the highest chances are of headache (around 43\%), second highest is of cough (around 32\%), and this trend goes on. Furthermore if you analyse $\varepsilon$ = 1, a clear majority of headache can be seen at 70\%, and cough being the second prominent with 25\%, however, the minor values such as dehydration and malaria are reduced to approximately 0\%. This means, maximum utility, but minimum privacy. One can even increase the value of $\varepsilon$ to more than 1 and there is no limit for highest value of $\varepsilon$. However, using a high value of $\varepsilon$ will reduce the privacy protection even further by increasing the utility~\cite{cutref01}.
}

\subsection{\cont{Contributions of This Survey Article}}

\cont{While some researches carried out research in surveying concept of privacy preservation in blockchain, however, to the best of our knowledge, no work that provide an in-depth discussion about integration of differential privacy in blockchain technology have been presented in the past. In this paper, we provide state-of-the-art literature from the perspective of differential privacy and its integration in modern blockchain technology. In conclusion, the key contributions of this article are mentioned as follows:}

\begin{itemize}

\item \cont{We provide detailed information about functioning and integration of differential privacy in blockchain from implementation perspective.}
\item \cont{We survey the current state of privacy in four major blockchain networks and highlight the possible use cases for integration of differential privacy in them.}
\item \cont{We provide an in-depth discussion about integration of differential privacy in each layer of blockchain model.}
\item \cont{We carried out detailed survey of all technical works that have been carried out regarding differential privacy and its integration with practical blockchain based applications.}
\item \cont{We provide detailed technical information about various blockchain based future applications that will require integration of differential privacy in them.}
\item \cont{We highlight current challenges, future directions, and prospective solutions for integration of differential privacy in blockchain.}

\end{itemize}

\section{\cont{Comparison with Related Survey Articles}}

\cont{In literature, a lot of work has been carried out in the field of privacy and blockchain. In order to draw a comparison between the presented article and the previous surveys, we develop selected 10 state-of-the-art and most relevant surveys from reputed journals and presented a comparison between these surveys and our work in Table~\ref{tab:comptable}.}\\
\cont{Starting from 2018, one of the pioneering survey article targeting security and privacy issues of Bitcoin was given by Conti~\textit{et al.} in~\cite{survey01}. The survey comprehensively highlighted all the underlying technologies of Bitcoin along with their feasibility and robustness analysis. Afterwards, the authors analysed security and privacy threats to these underlying technologies and drew a conclusion that significant attention is required in order to overcome all these challenges. In the similar timeframe, another survey article focusing over anonymity and privacy of Bitcoin-like crypto and digital currencies was published by Khalilov~\textit{et al.} ~\cite{survey02}. This survey presented a detailed investigation of various technologies that are being integrated with cryptocurrencies. After presenting some background literature, the article technically analysed methods of privacy and anonymity in Bitcoin and then mentioned research outcomes of these methods. Finally, the authors summarized their article by mentioning the guidelines designing and development of such platforms that can effectively preserve privacy of cryptocurrency user. In the quest of highlighting all possible privacy preserving solutions, a very detailed literature review has been carried out by authors in~\cite{survey03}. After carrying out extensive literature review, authors classified all literature from perspective of four data types named as identity data anonymization, key management data, on-chain data protection, and transaction data protection. Afterwards, authors discussed privacy preserving solutions for each datatype and categorized them further from integration perspective.} \\
\cont{Due to the secure cryptographic nature, the usage of blockchain is not just restricted to cryptocurrencies, instead, plenty of daily life applications are using blockchain. A detailed survey highlighting the security services that are being provided by using blockchain technology have presented by authors in~\cite{blchref04}. Authors presented a thorough summary of security services such as confidentiality, authentication, and privacy beside with providing complete technical details about integration of blockchain in these services. Another pioneering work in the field of blockchain and its applications have been published by Ali~\textit{et al.} in~\cite{newref02}. The authors covered almost all type of applications of blockchain in the domain of Internet of Things. The work first highlighted the use cases and integration of blockchain in IoT devices from perspective of various features and blockchain types. Afterwards, authors carried out an intensive literature review for integration of blockchain and IoT by highlighting challenges, issues, and currently running projects. \\
Moving back to privacy protection, an in-depth survey about privacy protection in decentralize blockchain systems have been presented by Feng~\textit{et al.} in~\cite{survey06}. The article first highlighted all the privacy requirement, and afterwards analysed all privacy issues in a very comprehensive manner. Moving further to next phase of article, authors discussed certain defence mechanisms for protecting privacy of blockchain by focusing mainly over cryptographic mechanisms such as ring signature, etc. From perspective of privacy preservation in IoT based blockchain, an extensive survey has been presented in~\cite{newref03}. The paper first carried out an extensive survey of all privacy preserving approaches of blockchain based IoT scenario from implementation and integration perspective. Furthermore, authors discussed future challenges and prospective research directions from an in-depth perspective. Another book chapter on privacy preserving techniques of blockchain have been published by Cui~\textit{et al.} in~\cite{survey08}. In the chapter, authors reviewed various typical cryptographic approaches that are being used in blockchain to preserve privacy of its users.} \\
\cont{A very unique article from point-of-view of attack surface in the domain of blockchain have been written by Saad~\textit{et al.} in~\cite{survey09}. The article systematically presented a detailed survey on attack surfaces, by having the major focus on public blockchain networks. The article first discussed the implication of various types of attacks in blockchain and then overviewed the vulnerability of various blockchain platforms toward these attacks. Afterwards, authors analysed various types of consensus mechanisms, forks, and then studied the effect of hash rate on orphaned blocks in forks. Finally, authors highlighted various application scenarios and types of attacks being carried out by adversaries on these applications. The last article in our list also focuses over attacks and security vulnerabilities of blockchain based IoT and IIoT systems and is presented by Sengupta~\textit{et al.} in~\cite{survey10}. The authors carried out a detailed investigation on security challenges along with their solutions in blockchain based IIoT domain. Authors first analysed cyberattacks for IIoT and then focused over ‘Tangle’, an IoT specific framework which serve as a basic structure for IoTA (also known as directed acyclic graph (DAG)), which was introduced by researchers to eliminate the need of miners in the network. Finally, the authors highlighted various research direction in the field of security and attacks in blockchain based IIoT domain.} \\ 
\cont{After analysing all this discussion, it can be concluded that plenty of survey articles are available, however, none of them address the topic of integration of differential privacy in blockchain from an in-depth perspective.
}\\


\begin{table*}[t!]
\begin{center}
 \centering
 \makegapedcells
 \footnotesize
 \captionsetup{labelsep=space}
 \captionsetup{justification=centering}
\caption{\textsc{\\ \cont{Privacy Preservation Status in Various Blockchain Platforms}}}
  \label{tab:platformtable}
 {\color{black} \begin{tabular}{|p{2cm}|p{4cm}|p{5cm}|P{4cm}|}
  	\hline
\rule{0pt}{3ex}
\bfseries \cont{Platform Name} & \bfseries \cont{Current Privacy Status} & \bfseries \cont{Possible Attacks} & \bfseries \cont{Possible Use cases of DP} \\
\hline

\multirow{3}{*}{}

\rule{0pt}{3ex}
\centering \bfseries \cont{Bitcoin} &  \cont{Pseudonymity (Hash of public key as receiver address)} & \tabitem \cont{Punitive and feather forking attack} \newline \tabitem \cont{Linking Attack} & \cont{Sender and receiver address protection }\\ 
\hline

\rule{0pt}{3ex}
\centering \bfseries \cont{Ethereum} &  \cont{Hash Functions} & \tabitem \cont{Deanonymization attack } \newline \tabitem \cont{Eclipse attack} & \cont{Protecting network-wide privacy} \\ 
\hline

\rule{0pt}{3ex}
\centering \bfseries \cont{Hyperledger Fabric} &  \cont{Pseudonymity via Symmetric Encryption} \& \cont{Hash Functions} & \tabitem \cont{Data analysing attack} & \cont{Preserving data during consensus} \\ 
\hline

\rule{0pt}{3ex}
\centering \bfseries \cont{IOTA} &  \cont{Quantum Resistant Hash based Signature} & \tabitem \cont{Trinity attack} \newline \tabitem \cont{Wallet exploitation attack} & \cont{Wallet address protection} \\ 
\hline

 \end{tabular}}
  \end{center}
\end{table*}


\section{Motivation Behind Usage of Differential Privacy in Blockchain} \label{dpblch}
Blockchain is a revolutionizing technology that has changed the concept of digital form of trading or data storing. Because of its decentralized nature, blockchain is considered to be next generation of secure storage. However, certain \tns{issues regarding blockchain still require solution before its implementation in everyday life scenarios.} One of the major parameter that requires considerable attention is preserving data and transaction privacy for blockchain applications. As identification of every user of blockchain in the decentralized network is carried by its public key, which means that the identities are not 100\% private or anonymous. \jpdc{Therefore, any adversary can act as a third-party analyst to analyse the transactions taking place inside the network and in turn may be able to infer original identities of individuals.\\ }

\subsection{\cont{Advantages of Integration of Differential Privacy in Blockchain }}
In this section we discuss some of the basic requirements of using privacy preservation strategy in blockchain and further provide details that how differential privacy outperforms all other privacy preserving strategies.

\begin{itemize}

\item \cont{If we analyse decentralized nature of blockchain,} various scenarios of blockchain can be observed that are not protected and needs more privacy indulgence to protect personal data of blockchain nodes. For example, whenever a transaction occurs in a financial blockchain system, the information about the transaction is broadcast throughout the decentralized network. This broadcast is done to ensure that every node has updated information, and the ledger that keeps the records is uniform throughout the network. \jpdc{However, this information can be maliciously used by any adversary to keep records of a specific individual and backtrack all his transactional and financial details. In order to protect privacy of this transaction, Laplace and Gaussian mechanisms of differential privacy can efficiently perturb specific values to ensure identity privacy~\cite{jpdcref18, chenref02}.}

\item \jpdc{Data stored in certain decentralized blockchain databases can also be utilized to conduct surveys~\cite{chenref03}. However, if the survey conducting organization becomes an adversary and tries to extract personal information, then the complete privacy of blockchain system can easily be compromised. In here, efficient Exponential query evaluation mechanism of differential privacy can play its part and protect private information of decentralized databases from such adversaries.} 
\item \jpdc{Till now, only anonymization and derivatives of anonymization strategy have been discussed in the literature to preserve individual privacy of blockchain~\cite{blchref08}. However, various experiments revealed that anonymization is not a complete form of privacy, as any anonymized data can be combined with similar datasets to reveal personal information~\cite{dpref05}. In order to overcome these issues, integration of differential privacy with modern blockchain technology can be a viable solution because of its dynamic nature and strong theoretical guarantee~\cite{jpdcref18}.} 
\item \jpdc{Dynamic nature of differential privacy makes its implementation suitable in blockchain scenarios. For example, in case of real-time data transmission or broadcast in blockchain applications, point-wise data perturbation strategy of differential privacy can efficiently add noise to data without disturbing the level of required accuracy~\cite{dpref06}. In point-wise data perturbation mechanism, first of all error rate is calculated and afterwards noise is calculated using that specific error rate. After computing specific noise value, the noise is added to the certain value to protect its privacy. \cont{Now the recorded value is differentially private, and any observer adversary cannot accurately guess the exact value or presence or absence of any individual inside decentralized database.}}
\item \jpdc{Before providing statistical blockchain databases to analysts for analysing, they can first be protected using differential privacy. \tns{In case of statistical blockchain data, a sense of indistinguishably can be created via differential privacy, and the query analyst could not predict with conviction regarding availability of a specific blockchain node in the dataset.} }
\item \jpdc{Another use case of differential privacy in blockchain technology could be the efficient preserving of identities of individuals \cont{during broadcast, in which differential privacy} can perturb the identity in such a way that the information is still useful to complete transaction, but the nodes or adversary in the network will not be able to judge the exact identity of sender or receiver.}  

\end{itemize}
\jpdc{Hence, the formal definition and theoretical model of differential privacy has the ability to control the privacy of only the crucial information within a set of data. Therefore, we can say that addition of differential privacy in blockchain-based applications can be proven fruitful in certain tremendous ways. A brief summary about integration of differential privacy in certain blockchain scenarios is given in Section~\ref{newsection}.}

\subsection{\cont{Current State of Privacy in Various Blockchain Platform}}

\cont{Blockchain has been widely used by researchers and developers to develop various frameworks according to the need and requirement. Blockchain platforms can be classified into various groups depending upon their permission requirements, application, coding flexibility, and transaction types~\cite{platref01}. Similarly, all platforms have their own privacy preservation style, some come up with basic cryptography, while others have quantum-resistant cryptography for anonymity. In this section, we discuss four major blockchain platforms, along with providing their current level of privacy, possible vulnerabilities, and how differential privacy will have effect after its integration in them. A detailed demonstration regarding all mentioned attributed is given in Table~\ref{tab:platformtable}.}
\subsubsection{\cont{Bitcoin}}
\cont{Bitcoin is considered to be the originator of blockchain, because blockchain technology came into limelight after the deployment of Bitcoin in 2008. Bitcoin as a currency has attracted attention of all banks and government because of having a market capitalization of billions of dollars~\cite{survey01}. However, from the perspective of platform, Bitcoin is a permissionless and public ledger that works over the phenomenon of solving cryptographic puzzle in order to mine the block. Bitcoin platform does not provide its users the facility to write any smart contract or to carry out operations other than currency trading (sometimes data storage), because the sole purpose of creation of Bitcoin was to develop a decentralized payment system which will be free from any intermediaries. Therefore, the major focus of Bitcoin was and is to ensure and provide maximum security to its users. No doubt, because of complex proof of work (PoW) puzzle, cryptographic hashing, and salient security features, Bitcoin is one of the most secure cryptocurrency. Although, it still lacks in various fields from perspective of privacy preservation.}\\
\cont{Bitcoin being a publicly available network is vulnerable to many privacy threats, and linking attack is one of the most prominent one. In linking attack, the public addresses of various transactions are compared with each other in order to trace the person behind these transactions. As the hash of public key is used as a receiver address in the bitcoin transaction, therefore, it can be traced back to its original owner. Similarly, punitive \& feather forking attack is also carried out by certain attackers to blacklist the transactions originating from a specific address~\cite{survey02}. In order to mitigate this attack differential privacy can serve as a handy tool, because dynamic nature of differential privacy can protect sender and receiver addresses by effectively anonymizing the addresses by adding adequate amount of random noise during transaction process.  Furthermore, the public ledger can also be protected from linkage attacks by randomizing the transaction addresses using differential privacy, so that there is some degree of noise in the values reported to ledger. This can be carried out by mining peers or the transaction nodes. Furthermore, in this way plenty of privacy attacks related to Bitcoin addresses can be controlled in an efficient manner.}
\subsubsection{\cont{Ethereum}}
\cont{Ethereum is also an open source platform, which was first introduced as a competitor of Bitcoin cryptocurrency. However, the user-friendly development environment and decentralized applications (DApps) functionality of Ethereum made it among one of the most favourite platforms of blockchain developers. Ethereum has the capability to run smart contract and a vast number of decentralized applications can be developed using this feature ranging from a basic tic-tac toe game to a complex energy trading system. In order to execute a smart contract, one needs to host an Ethereum virtual machine (EVM) at every node~\cite{platref02}.}\\
\cont{From privacy perspective, Ethereum provides cryptographic hash functions as mean of privacy, and transaction and other records are protected via using cryptographic mechanisms-based privacy~\cite{platref03}. However, addition of these cryptographic mechanisms does not guarantee complete privacy, because Ethereum being a public platform allow its users to see the decentralized ledger. Therefore, the most prominent privacy attack on Ethereum is deanonymization attack, in which data from distributed ledger is deanonymized by carrying out linking and tracing the features with other databases. In order to resolve this issue, differential privacy provide the functionality of addition of noise in the stored distributed ledger records. The randomness and noise of differential privacy can be used in multiple ways, for example one case could be to protect the complete ledger by adding random noise for non-trusted users, or for users that do not have a specific stake in the network. Another use-case could be to only allow query evaluation in public ledger to analyse any record or previous transaction, and during this query evaluation, a noise could be added to protect privacy. Similarly, the smart contract of Ethereum do also provides its developers the functionality to add differential privacy in its truncations, which can also be a use-case, and for generalist users, a basic DApp can be developed to control the noise controlling parameter ($varepsilon$) according to their need.}

\subsubsection{\cont{Hyperledger Fabric}}
\cont{Hyperledger can be termed as a sub-project which comes under umbrella of a very large open-source project of Linux Foundation, which was created to integrate blockchain in industrial domain~\cite{platref04}. Around 270 organizations have joined Hyperledger community till now, and many industries are considering joining this project because of its enormous advantages~\cite{platref05, platref01}. Nevertheless, Hyperledger being an open-source development platform provides its users all customizable functionalities, from developer and layman users perspective. For example, Hyperledger Fabric provides layer-based architecture to its developers via which they can modify Hyperledger according to their needs, and on the other hand it provides a very handy and user-friendly user interface for the buyers or organization personals.}\\
\cont{In order to protect privacy of its organizations, Hyperledger Fabric provides pseudonymity via symmetric encryption and hash function. Since, Hyperledger Fabric is a permissioned blockchain network, therefore, it can control ‘who to see the ledger’, however, this does not completely guarantee privacy. Because some participating members can be compromises and complete  data can be attacked via data analysing attack, in which various machine learning algorithms are used to infer into privacy of participants. Private data of customers, their bank details, and similar other features can be extracted if some adversary gets access of protected data. But, in order to ensure trust in the network, it is compulsory to record every change to the ledger. The privacy of customers and organizations can be protected by using differential privacy in Hyperledger. For example, consensus in Hyperledger Fabric is carried out by authoritative nodes (or the organization nodes). These nodes have the capability to modify the smart contract before deploying. Therefore, these nodes can effectively protect certain parameters of smart contracts in order to add differentially private noise in them. In Hyperledger Fabric, all the inter-operations can be protected in different ways via developing smart contracts accordingly.}

\subsubsection{\cont{IOTA}}
\cont{IOTA can be termed as a publicly distributed ledger that works over the phenomenon  of directed acyclic graph (DAG). In IOTA, there are no miners, no chain, and not even blocks, however, because of its publicly available distributed ledger, it is considered a part of blockchain family~\cite{platref06}. In IOTA DAG, each transaction/site always has a direction towards another transaction, this is how these transactions are linked with each other and these directions help traverse back to any required transaction. Each site also has connection to at least two other sites, which are called edges and these connections are for the purpose of validation of that specific site. In order to add a transaction, the algorithm selects two tips of the graph randomly to add the next one. The next added transaction verifies the previous two transaction and becomes the part of the directed acyclic graph. This process means that every new transaction does also confirms two other transaction, which increases scalability and robustness of the network. Accumulative weights are added in each transaction with respect to the weight of it edge transaction, this accumulative weight is then further used to verify the complete graph from any end.} \\
\cont{Furthermore, IOTA uses quantum resistant hash-based signature to enhance its security and privacy~\cite{platref06}. However, despite of this technique, IOTA is still vulnerable to certain wallet theft attacks, and one of the most prominent one recently happened on 12 February 2020, which is also known as “Trinity Attack”~\cite{platref07}. In this attack, IOTA users lost their funds to an unknown attacker, which carried out a seed-based attack from wallet public keys. This was certainly a challenging incident for IOTA developers; however, this could have been controlled by using differential privacy during seed generation process for wallet keys. For example, the online generators which were used to generate seed for wallet keys could have protected the privacy a bit further by randomizing the keys via differential privacy-based randomness in a way that attacker will not be able to link itself to the exact private data and might will not be able to get into accounts of IOTA users. This indeed is a long journey, but we believe that dynamic nature of differential privacy protection could be an important step to protect privacy of such blockchains.
}\\


\begin{figure*}[htp]        
\centering
\includegraphics[scale = 0.8]{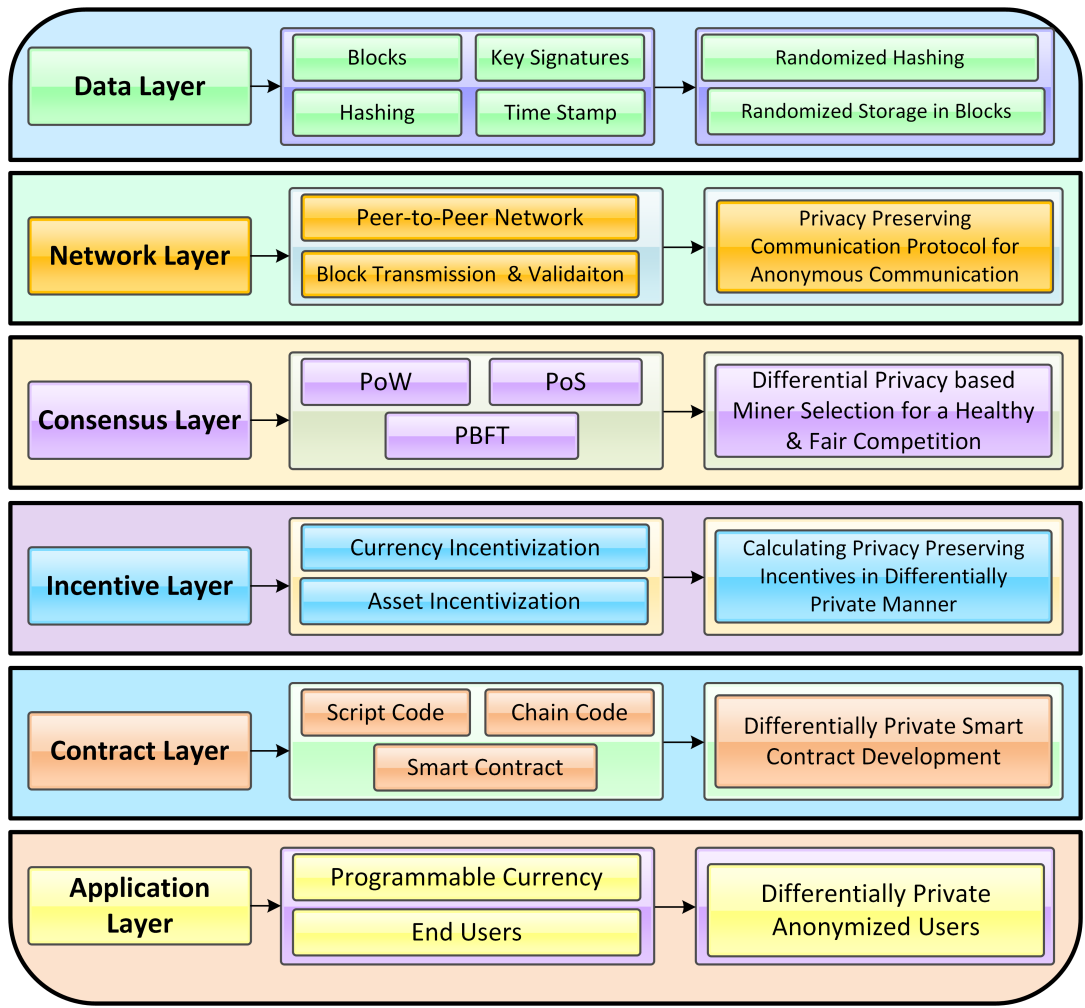}
 \caption{\cont{Integration of differential privacy in blockchain layer based architecture. Dynamic noise addition mechanism of differential privacy can integrated in every layer of blockchain by adjusting the noise to an optimal value.}}      
  \label{fig:layerfigure}   
\end{figure*}


\section{\cont{Integrating Differential Privacy in Each Layer of Blockchain Technology}}

\cont{Blockchain indeed is a publicly distributed ledger, however, from in-depth perspective its architecture has further been classified into six different layers by researchers~\cite{layer01, layer02, layer03, layer04}. Each layer has its own functioning and privacy requirements, for example the privacy requirements at user end will be totally different from privacy requirements while formation of blocks in data layer or while carrying out consensus in consensus layer. This section formally discusses the functioning, privacy requirements, and need and integration of differential privacy in each layer of blockchain architecture. A detailed figure containing all architectural layers along with differential privacy integration have been given in Fig.~\ref{fig:layerfigure}.}\\

\subsection{\cont{Data Layer}}
\cont{Data layer is the core layer of blockchain, which plays the most important role in the design and development of decentralized distributed ledger. This layer comprises of blocks, which are storage entities consisting of cryptographic signatures, hashing, timestamp, data, etc~\cite{layer05}. These blocks are chained together to form a chain like structure, and all this is happening within the vicinity of data layer. A typical block inside the data layer can further be classified into two parts: the header, and the body. Header store the information related to formation of chain structure such as metadata, previous block hash, current block hash, timestamp, Merkle root value, nonce, etc. On the other hand, the body comprises of transaction, or any other type of storage that is the required in formation of block, such as chaincode, etc. Different blockchain applications store, manage, and organize blocks in a different manner, which purely depend upon their application scenario. \cont{For example, the block size of Bitcoin is 1MB, which is only used for transaction, however, the block size of Hyperledger Fabric is reconfigurable and it purely depends upon the type of organization using it up, and it can even be more than 1GB~\cite{layer06}.}\\
From perspective of privacy preservation, the complete chain of blocks can be downloaded and analysed to intrude into private data stored in the blockchain. Therefore, certain blockchains try to integrate some privacy preserving medium in every step and aspect of this layer. For example, different type of hashing mechanism such as SHA-256, MD-5, SHA-1, etc. Similarly, public and private keys of users are protected via cryptographic signatures, etc. We believe that integration of differential privacy can aid and increase privacy in certain core components of this layer. }

\subsubsection{\cont{Hashing}}
\cont{Hashing is the fundamental aspect that ensures immutability of blockchain, because hash is a parameter that is generated via specific hashing function and is always a unique expression. For example, even if I change a dot ‘.’ in the data of block, it will generate a completely new hash. Hashing is used to link one block to its previous block, as header of each block do consist of hash of current and previous block, and this chain is formed and the blocks in blockchain are traversed back using these stored hashes. Hashing in blockchain is carried out via certain strong hashing mechanisms such as SHA-256, SHA-1, etc~\cite{layer07}. No doubt, these hashing functions are strong and some of them are even resilient to quantum-based attacks as well. However, these security aspects do not guarantee the privacy of data via which these hashes are calculated. For example, for the same input data the output hash will always remain the same, and in case of some intruder gets access to some limited data inside a block and wants to predict remaining data then multiple computation of hashes by varying data will finally provide the link to exactly similar data  and the adversary can predict with confidence about each and every attribute of the block.} \\
\cont{In order to overcome this issue researches needs to be carried out from perspective of integration of differential privacy-based randomization in hashes. Some works from apple, discussed the integration of differential privacy in during hash calculation to ensure the privacy of hash functions~\cite{layer08}. However, not enough literature is available over this domain, therefore there is a need to carry out extensive research. According to our point-of-view, the randomness produced via differential privacy can be considered as a key aspect toward protecting hash functions privacy.}

\subsubsection{\cont{Data Blocks}}
\cont{Blocks are the core entities of blockchain as they store all transactional and storage records in a secure manner. As discussed above, blocks can be divided into two major components: header and body. Regarding header privacy, we have discussed about in the above subsection about private differential privacy based hashing. Therefore, in this section, we discuss about need and integration of differential privacy in body of a block. Transactions and all other data are stored in the body of block in a specific data structure called as Merkle tree. There can be multiple transactions, multiple files, and even multiple smart contracts inside the body of block. From privacy preservation perspective, this is the most critical parameter. As getting access to the body of block means getting access to almost all components of blockchain. Therefore, protecting privacy of block body is one of the most important aspect in the sight of all blockchain privacy researchers.}\\
\cont{To make blockchain more private, and to ensure privacy in the blockchain, plenty of works are being carried out. For example, few works highlighted to use only hashes of data instead of complete data, some works highlighted to apply anonymization while recording data to blockchain, some works use cryptographic methods to protect privacy of blockchain, and certain works used differential privacy to protect such data. A detailed discussion about protecting blockchain data mainly within data blocks via privacy preservation strategies have been given in Section~\ref{newsection}.} \\ 
\cont{Since plenty of works have been done in this field of integration of privacy in blockchain, we believe that integration of differential privacy in blockchain body before recording data to blockchain ledger is the most optimal way. Differential privacy provides its users with the leverage to control the amount of privacy, along with this it also provides its users the dynamic functionalities that the users can add privacy according to the sensitivity of data. Since in data blocks, the transactions and chain-code are recorded, therefore, it is important to introduce a certain level of randomness, which is best possible via differential privacy. For example, if a decentralized energy auction is carried out on blockchain network, then protecting bid privacy is the most important thing. However, if one records all truthful values on blockchain, then an adversary can easily trace back to original bids. Therefore, in order to protect such privacy, data is protected via adding randomness in the data. The addition of this randomness does not affect the overall result of auction; however, it ensures that the recorded bids do not leak user privacy.}

\subsection{\cont{Network Layer}}

\cont{In blockchain, network layers play an important role of timely dispersion and dissemination of message via efficient communication protocols. The aim of this layer is to disseminate the data and blocks generated from the above data layer to all the participating/authorized nodes~\cite{layer09}. The data messages could be of different types and can be for different participants depending upon the requirement, for instance, in case of permissioned blockchain, the data regarding consensus is only disseminated to authority nodes, however, in case of public blockchain, all nodes can participate in consensus and can propagate the block in case they cope-up with the requirements. As it can be seen that network layer is all about communication, therefore, making this communication secure and private is the most important aspect of security researchers working in this field.} \\ 
\cont{To protect security of this layer, researchers integrated plenty of secure communication methods with blockchain that ensure that none of the communication gets leaked. However, the area of privacy preserving communication still lack as compared to advances in security aspect. According to our analysis, integration of differential privacy in such communication protocols can effectively protect privacy of such protocols.}

\subsubsection{\cont{P2P Network-wide Communication}}
\cont{The peer-to-peer (P2P) network in blockchain establishes a connection between all the participating nodes in the most efficient manner. As in majority of blockchain scenarios, there is no central entity, and any node can join and leave the network at any time, and similarly, all nodes can contact and reach other nodes without any limitation. Therefore, a communication medium that can provide such advantages without having a node failure is required. In order to do so, plenty of researches are being carried out in the field. For example, Kan~\textit{et al.} proposed a broadcast propagation protocol by using tree routing mechanism~\cite{layer10}. Similarly, Jin~\textit{et al.} proposed a named data networking based communication protocol for Bitcoin network operating over phenomenon of blockchain~\cite{layer11}. Similar to this, plenty of other works have been carried out from researchers in this field, which uses various techniques to provide efficient and secure communication.} \\ 
\cont{No doubt, researchers are actively working over the enhancement of integration of security and efficiency in blockchain communication, however, the field of privacy preserving communication still has plenty of gaps which are yet to be explored. For example, Song~\textit{et al.} highlighted that an adversary can eavesdrop the network and can potentially attack privacy if the communication is not protected via some privacy preserving mechanism~\cite{layer12}. However, to the best of our knowledge, no work that specifically integrates privacy in communication network of blockchain have been carried out yet. Therefore, there is a need to carry out research in this domain in order to propose such mechanisms. According to us, the integration of differential privacy in the communication can be an important step ahead. For instance, protecting the sender and receiver private information by addition of randomness while block propagation and broadcasting, and protecting the addresses via adding some noise before transmitting messages across the network can ensure privacy in such P2P networks. }

\subsubsection{\cont{Block  Transmission \& Validation}}

\cont{Transmission and validation of block serves the major purpose of recording of block on the distributed ledger. Whenever a miner mines a block in the blockchain, this block is sent to all participating peers in order to get validated. The peers then validate the block, by verifying the data inside the block by comparing hash values, and once the block gets validated, it is then recorded on tamper-proof ledger~\cite{layer13}. This complete step-by-step process of recording of block on the ledger do require approval and verification at approximately each stage, which require sharing of private information up to some extent, for example ‘miner ID’, ‘transmitter ID’, ‘validator ID’, etc. These IDs are cryptographically protected by using encryption based techniques, however, recent advances have highlighted that quantum based attacks can easily overcome these protection methods and can infer into private information of participating nodes.} \\
\cont{Therefore, integration of a privacy protection mechanism during block transmission and validation is important. According to our point of view, randomization mechanism of differential privacy can be a viable step towards protecting this privacy. For instance, differential privacy can add randomness in the received block in a way that the receiver is not sure whether he received the block from ‘X’ participant or ‘Y’ participant. However, the transmission protocol or the developers can distinguish between it, so that in case if someone cheats, then he/she could be caught. In this way, adversary will not be able to infer into privacy of participants even if they get access to the sender address, because the adversary could not predict with conviction about the presence or absence of any particular sender. The same can be done during block validation process, for example, a specific participant ‘Z’ validated this specific block or not, this information can be protected by adding some degree of randomization in the complete process.} \\

\subsection{\cont{Consensus Layer}}
\cont{As the name suggests, consensus layer comprises of a consensus algorithm which is used to reach to a single point of agreement among all untrusted nodes in distributed decentralized blockchain environment~\cite{layer14}. Consensus algorithms vary with various blockchain networks and types, for example, in a permissioned network, consensus is only carried out between authoritative nodes and only permissioned nodes can take part in consensus. However, in a public network, everyone can take part in the consensus and can mine block after fulfilling the requirements, such as Bitcoin network~\cite{layer15}. In this section, we discuss some famous consensus variants and provide the need and way to integrate privacy in them.}
\subsubsection{\cont{Proof of Work}}
\cont{Proof of work (PoW) which was originally proposes by Nakamoto in 2008, served as a progenitor of all other blockchain protocols~\cite{layer16}. Nakamoto proposed the idea an incentive based consensus algorithm for a permisionless environment, in which miners can compete with each other in solving cryptographic puzzle, which was also referred as cryptographic block-discovery game by some researchers. In this consensus, miners accumulate all transactions from mining pool and try to calculate a hash value from these transactions via secure hashing algorithm. If the computed hash is less than the target hash, then the miner gets the permission to mine the block in the network, and in return the miner gets the designated mining reward. The computed hash along with the transactions and time-stamp is then sent to the validating nodes as mentioned in the above layer.}\\
\cont{As from the mechanism, it can be seen that miners are actively solving the cryptographic puzzle in order to mine the block, so this develop a competition among miners to mine the block as soon as possible in order to win the mining reward. This indeed is a healthy competition because miners use their resources to win the race, but on the other hand this also raises plenty of privacy vulnerabilities. For example, a miner which is actively solving the puzzle and mining the blocks because of its computation power can be caught in the sight of miners which cannot compete him in the computational powers, and then they can go for unethical means to harm that miner or its computational power, for example, launching of attacks to a specific miner, such as punitive and feather forking attack discussed above in which transactions form a particular node or miner gets banned. Therefore, it is important to introduce a specific degree of randomness in the PoW mining. This can be done by integration of differential privacy in the miner selection mechanism for PoW, for instance, for let’s say 20\% of times the miner such miner is chosen which solves the block but does not get to the target value. This will reduce the sense of rivalry among the miners and in this way, miners having low computational power will also be able to compete and win the mining reward.}
\subsubsection{\cont{Proof of Stake}}
\cont{Proof of Stake (PoS) got popularity when Ethereum cryptocurrency first introduced  this algorithm as an alternative to PoW in order to overcome the issue of using extensive computational power~\cite{layer18}. In PoS consensus algorithm, a miner is chosen on the basis of its stake inside the blockchain network. In order to have that stake, the miner have to deposit specific number of coins/tokens in the network, and this stake will be taken by network in case if miner behaves maliciously. However, the major issue with this basic idea was that miners used to deposit the stake when they wanted to mine, and elsewise they used to remove the stake from the network. In order to overcome this issue, researchers introduced the concept of coin age, which means the oldest the coin is, the more it will be contributing in the stake~\cite{layer17}. This idea gave a new dimension to PoS consensus algorithm, and afterwards it has been widely applied to plenty of domains.} \\
\cont{Overall, this concept of stake based mining is quite appealing, however, it may pose serious threats to participating miners in case if their private information gets leaked. For example, if all the participants of blockchain networks gets to know about a person that person ‘X’ has 40\% of stakes in the network, then ‘X’ could have plenty of potential threats, such as theft threat, blackmailing, etc. Therefore, it is important to protect privacy of such individuals. Currently, the cryptography based privacy does not comply with all the privacy requirements and there is a possibility of such privacy leakage. That is why, integration of differential privacy can be a viable solution. For example, instead of always choosing the miner on the basic of stake, a miner can randomly be chosen via Exponential mechanism of differential privacy, which will introduce a sense of randomization in the network, and nobody will be targeting any specific individual and will think of moving further to carry out a healthy competition. Therefore, researches to integrated differential privacy in PoS should be carried out.}
\subsubsection{\cont{Practical Byzantine Fault-Tolerant (pBFT)}}
\cont{Practical Byzantine fault-tolerant (pBFT) was introduced to carry out consensus in a permissioned environment where only authoritative nodes such as various industries take part in the consensus. These types of consensus mechanisms are different from other traditional mechanisms, and they were introduced to consider Byzantine fault tolerance (BFT) in the blockchain network, which assumes that nodes can go through network fault, downtime, and similar other nonhuman issues. However, malicious nodes are excluded from this environment, because they are continuously trying to find loopholes and faults in the network~\cite{layer19, layer20}. Therefore, the goal of pBFT algorithm is to reduce the influence and effect of these adversary nodes even in the case of node failures. In order to function pBFT algorithm it is also important to assume that no more than one-third of the nodes in the network are malicious. For example, in order to mine a block two-third majority is required, and even if the remaining one-third becomes adversary, still, the blockchain will be able to function without having significant issues~\cite{layer04}.} \\
\cont{In order to execute the consensus, a leader is chosen, and the leader can take the decision of mining or rejecting the block once it gets request from the client side regarding invoking of service operation. Similarly, a leader is chosen on the basis of elections among all the authoritative nodes (e.g., among all collaborating industries). And in case if the leader behaves maliciously, it can be removed and penalized accordingly. Since it can be seen from the process that choosing the leader is an election based process, and since all the authoritative nodes are considered trusted, anyone among them can become the leader. However, this leader choosing phenomenon can raise a competition among the mining nodes and they can behave maliciously to keep on choosing a selective leader again and again by voting. On the other side, leader oriented targeted attacks can also be carried out on a specific leader if it gets chosen quite often. Therefore, it is a need to integrate privacy during the process of this leader selection. We believe that merger of differential privacy and pBFT algorithm can be a good combination, as differential privacy can randomize the complete process of leader selection. Furthermore, differential privacy will also provide the opportunity for miners to be leaders that can never become leaders because they do not have enough support from other miners. Therefore, there is a need to integrate differential privacy based randomness in pBFT consensus.}

\subsection{\cont{Incentive Layer}}
\cont{Incentive layer serves as a major force because it motivates all miners and participants to take part in development and functioning of blockchain by providing them incentives on the basis of their participation~\cite{layer21}. Incentives could be of different types, for example, Bitcoin and Ethereum provides incentives in the form of their respective coins, while some other blockchain networks provide incentives in the form of tokens that can be redeemed afterwards. In our discussion, we categorize incentives into two major categories; currency based incentives and asset based incentives. Currency based incentives usually work for cryptocurrencies or for the applications that deal with tokens. Similarly, asset based incentives could be for such platforms that do not deal with currencies or tokens such as Hyperledger Fabric~\cite{layer22}. These incentives can be provided to participants on the basis of their behaviour, for example a miner can get incentives if he/she able to mine block successfully. Similarly, a participant can be incentivized for behaving truthfully multiple times (let’s say ‘x’ number of times). Similar to this, point-based incentives can be given after winning a blockchain based game. Since the money is involved, this layer needs an additional privacy in order protect specific individual privacy. In this section, we discuss the integration of differential privacy in incentive distribution.}
\subsubsection{\cont{Differentially Private Incentives}}
\cont{Since incentive layer involves direct dealing with money and assets, therefore ensuring its privacy is one of the most critical aspect, and in order to do so, differential privacy is one of the most viable technique because of its adaptability. Differential privacy can be integrated and adjusted in almost all scenarios involving money trading, etc. It will not be wrong to say that integration of differential privacy in incentive layer is the most prominent one among all the works that have been carried out in integration of differential privacy in blockchain. For example, a differential privacy based auction mechanism for incentive layer have been proposed in~\cite{jpdcref01}. In the proposed auction mechanism, incentives and pricing values are calculated via differentially private manner in order to protect bid privacy. Similarly, whom to get incentive, whom to choose as winner for incentive, whom to choose for second winner, etc, all these questions can easily be answered by differential privacy mechanism in the most secure and private manner. Therefore, according to our point of view, differential privacy should be used while calculating incentives during any type of decentralized trading.}

\subsection{\cont{Contract Layer}}
\cont{Contract layer plays the role of backbone in modern blockchain applications, as it provides developers with the flexibility to develop their  require network~\cite{layer23}. Various types of scripts, codes, smart contract, and algorithms are used and deployed in this layer that enable the integration of complex transactions and functions into decentralized blockchain network. There could be different names for different types of scripts of this layer, for example, Ethereum call this layer ‘smart contract, Hyperledger Fabric call this layer “Chaincode”, although, the overall functioning of all these are same. This layer can also be termed as the layer containing all set of rules and logics for functioning. Furthermore, during deployment of a smart contact all the conditions and terms need to be met by all the participants for successful deployment. For example, a smart contract ‘X’ states that when ‘Joy’ will get 18 years of age, he will then get \$1000 from his dad’s saving. As it can be seen that the condition of participant getting 18 years old has been affiliated with this contract. So, the funds from savings cannot be deducted unless the participant ‘Joy’ reaches the specific mentioned age. Therefore, fulfilment of all the terms mentioned inside the blockchain is compulsory to be fulfilled . However, once the conditions got met and the deployment of smart contract gets triggered, then the execution of smart contract will carry out independently according to the written rules and it is not possible to stop this deployment.} \\
\cont{Similarly, another important aspect that needs to be taken care of while developing and deploying smart contract is the protection of metadata inside~\cite{meta01}. For example, smart contract is the entity that will be dealing with all the participants via pointers to their stored information, their linked accounts, etc. Therefore, a smart contract cannot blindly be deployed without making sure that none of the metadata information inside the smart contract paves the paths towards leakage of sensitive information.}\\ 
\cont{From the above discussion, it can be concluded that the deployment of smart contracts is a crucial step and it cannot be stopped once it starts execution, therefore, it is compulsory to ensure all security and privacy requirements of information and metadata inside it before starting the deployment. Private smart contract for blockchain is a vast field, however, all works carried out in this field are somehow related to other layers of blockchain. For example, if one integrates privacy to randomize user identities in transactions, then it gets directly linked with data layer. If one provides a smart contract for providing private incentives, then it gets directly linked with incentive layer. This is because the contract layer has direct relation with every other layer because it serves as a backbone of all other functionalities. Therefore, in this section we discuss some famous private smart contract and provide information regarding integration of differential privacy in contract layer.}\\
\cont{From viewpoint of current research, certain works highlighted the need of integration of privacy in smart contract and also proposed some mechanisms, such as Shadoweth~\cite{layer24}, Arbitrum~\cite{layer25}, Hawk~\cite{layer23}, Raziel~\cite{layer26}, etc. These works used various mechanisms to protect privacy such as some mechanisms use cryptography based techniques, some works used anonymization based mechanisms, and some works carried out enhancement of private key generation functions. Despite of such technological advancement, a full fledge work that targets the integration of differential privacy specifically in contract layer is still lacking in research. Randomness of differential privacy incorporation with flexibility of modern smart contract can be an important step towards development of privacy friendly blockchain. Similarly, due to lightweight nature of differential privacy, it can be incorporated at each layer of blockchain by development of an efficient smart contract. Therefore, such researches should be carried out to support all privacy requirements of blockchain.} \\

\subsection{\cont{Application Layer}}
\cont{Application layer is the top layer in the architecture of blockchain, this layer aggregates all data in an environment and interacts with the end user. This layer comprises of business oriented and client oriented applications such as user interface, digital identity, market security, IoT, intellectual property, and so on~\cite{layer27}. Similarly, application layer provides users a platform to carry out efficient and secure distributed and decentralized management. Furthermore, application layer also links various security and privacy features such as user authentication and data protection to user end. Therefore, application layer is most vulnerable to certain security and privacy threats such as ‘malware’ and reliability  attacks, which can directly affect performance of nodes and the whole blockchain network~\cite{layer28}. For example, DAO program on Ethereum was breached, which resulted in a loss of \$50 million worth of Ethereum tokens ‘Ether’~\cite{layer29}. Which further lead to formation of a hard fork on Ethereum network.}\\ 
\subsubsection{\cont{Differential Privacy in Application Layer}}
\cont{Differential privacy can play an important role in providing significant privacy to application layer of blockchain by ensuring the randomness in the network. For example, differential privacy can reduce the risk of attack by providing the randomized data on application layer instead of providing all the accurate data regarding other transactions and users. Differential privacy can be added in multiple manners in application layer, for instance, if an adversary wants to carry out any attack by analysing transactions of a public blockchain network, then, he (adversary) will not be able to predict with confidence regarding presence or absence of a particular participant in a block or a group of transaction because of added randomness via differential privacy. Similarly, differential privacy can also protect currency related privacy by randomizing any type of query evaluation carried out on data of public blockchain. For example, if an adversary asks a query to blockchain network that which user has highest number of tokens, from the private output results the adversary will not be able to predict with confidence that the output result if 100\% correct or not. In this way, the data of blockchain can also be used for plenty of statistical analysis without the risk of losing the private information. Therefore, we believe that integration of differential privacy will bring confidence among participating users.}

\subsection{\cont{Summary and Lessons Learnt}}
\cont{Architecture of blockchain is divided into six different layers named as data layer, network layer, consensus layer, incentive layer, contract layer, and application layer. This multi-layer architecture has been developed by researchers to understand the functioning of blockchain in detail from technical perspective. Each layer has its own functionalities and vulnerabilities, for example the attacks by adversaries on consensus layer will be totally different from application layer. Therefore, in order to understand the privacy requirement and integration of differential privacy in each layer we provide a detailed analysis of each layer.}\\
\cont{Starting from data layer, this layer comprises of the most important storage elements of blockchain such as blocks, hashes, etc. This layer is vulnerable to strong data linkage and de-anonymization attacks; therefore, randomization phenomenon of differential privacy helps prevent these attacks. The next layer is network layer which handles all communication of network, and in order to preserve privacy of this layer, differentially private communication can be integrated. Furthermore, the next layer named as consensus layer provide one of the most important functionality of blockchain; which is agreeing of non-trusted parties on a single point. Plenty of consensus algorithms can be used in this layer, however, protecting miner privacy is still considered the most important priority and differential privacy can successfully preserve that.\\
The next layer in the blockchain architecture directly deals with money/tokens and is named as incentive layer. Protecting privacy of participants at this layer is important because money is directly involved in this layer. In order to protect privacy at this layer noise addition mechanism of differential privacy can play a vital role. The next layer is contract layer in which coding is done; and it is considered the backbone of blockchain, because all functionalities and mechanisms depend upon the code being deployed by this layer. Therefore, protecting privacy by writing differentially private smart contracts will be an important step forward in research. Finally, the top layer is application layer which has direct link at user end. This layer has certain vulnerabilities and privacy threats such as data analysis attack, which can be encountered by differential privacy mechanisms.}


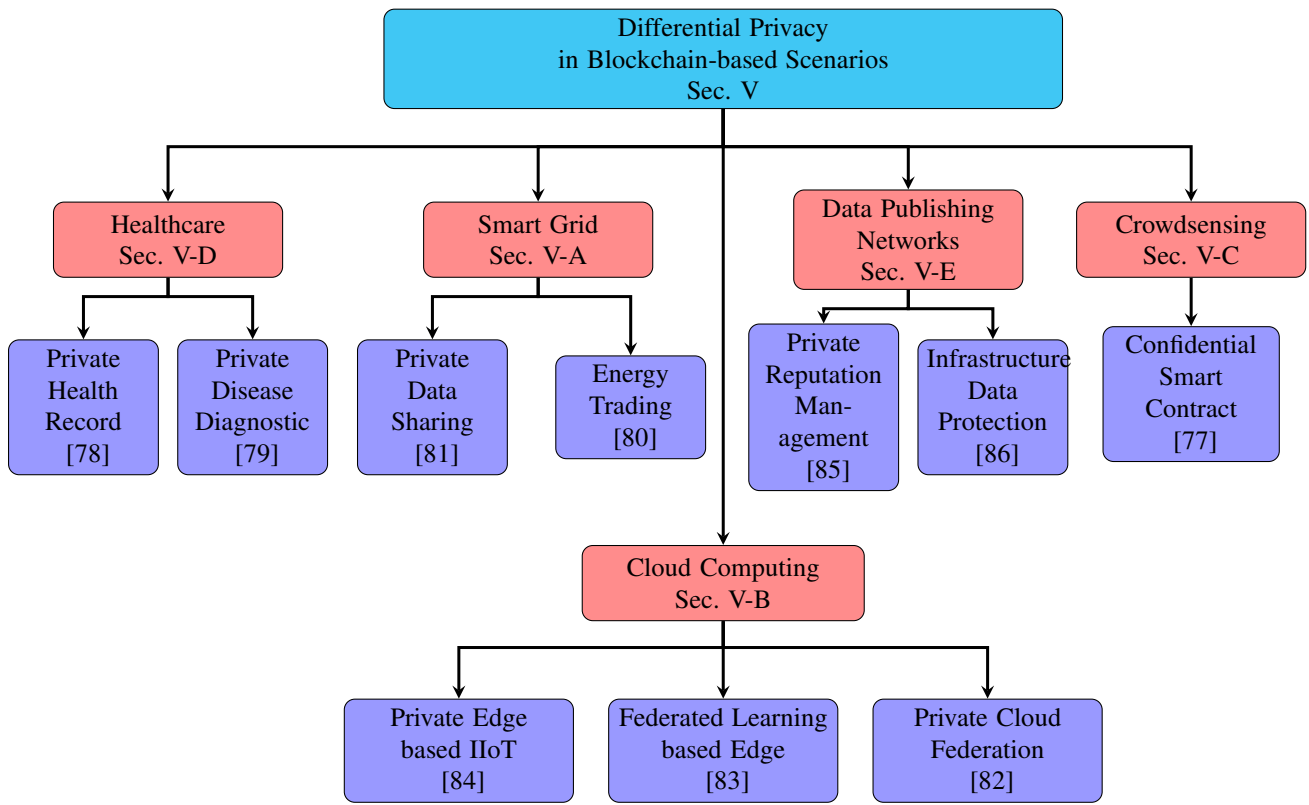
\begin{figure*}[]
\centering

\begin{tikzpicture}

\node [block,  text centered, fill=cyan!60, minimum width = 25em,  text width=25em] (a1) {Differential Privacy \\ in Blockchain-based Scenarios\\ Sec.~\ref{newsection}};



\node[block, below of=a1, yshift=-4em, xshift = 17.7em, text width=8em](b1){Crowdsensing\\ Sec.~\ref{crowd}};

\node [medblock, below of=b1, yshift=-3em, text width=6em] (b1c1) {Confidential Smart Contract\\ ~\cite{tnseref07}};

\node[block, below of=a1, yshift=-4em, xshift = -21em, text width=8em](x1){Healthcare\\ Sec.~\ref{health}};
\node [medblock, below of=x1, yshift=-3.5em, xshift = -3.2em , text width=5em ] (b1c2) {Private \\ Health Record \\ ~\cite{tnseref08}};
\node [medblock, below of=x1, yshift=-3.5em, xshift = 3.2em, text width=5em] (b1c3) {Private \\ Disease Diagnostic\\ ~\cite{techref02}};

\node[block, below of=a1, yshift=-4em, xshift = -7em, text width=8em ](n1){Smart Grid\\ Sec.~\ref{grid}};

\node[medblock, below of=n1, yshift=-3.5em,xshift = 3.5em, text width=5em  ](n1c2){Energy Trading\\ ~\cite{tnseref03}};

\node[medblock, below of=n1, yshift=-3.5em,xshift = -4em, text width=5em  ](n1c1){Private Data Sharing\\ ~\cite{tnseref04}};

\node[block, below of=a1, yshift=-17em, text width=10em ](m1){Cloud Computing\\ Sec.~\ref{cloud}};

\node[medblock, below of=m1, yshift=-3.5em,xshift = 10em, text width=8em  ](m1c1){Private Cloud Federation\\ ~\cite{tnseref05}};

\node[medblock, below of=m1, yshift=-3.5em, text width=8em  ](m1c2){Federated Learning based Edge\\ ~\cite{tnseref06}};

\node[medblock, below of=m1, yshift=-3.5em,xshift = -10em, text width=8em  ](m1c3){Private Edge based IIoT\\ ~\cite{techref01}};

\node[block, below of=a1, yshift=-4em, xshift = 7em, text width=8em ](b3){\cont{Data Publishing Networks}\\ Sec.~\ref{misc}};
\node[medblock, below of=b3, yshift=-3.5em, xshift = -3.2em , text width=5em ](b3c1){Private Reputation Management\\ ~\cite{tnseref09}};
\node[medblock, below of=b3, yshift=-3.5em, xshift = 3.2em, text width=5em  ](b3c2){Infrastructure Data Protection\\ ~\cite{tnseref10}};

\path [line] (a1)-- ($(a1.south)+(0,-0.5)$)-|(b1);
\path [line] (a1)-- ($(a1.south)+(0,-0.5)$) -|(b3);
\path [line] (a1)-- ($(a1.south)+(0,-0.5)$) -|(n1);
\path [line] (a1)--(m1);
\path [line] (a1)-- ($(a1.south)+(0,-0.5)$)-|(x1);

\path [line] (b1)-- ($(b1.south)+(0,-0.25)$) -|(b1c1);
\path [line] (x1)-- ($(x1.south)+(0,-0.25)$) -|(b1c2);
\path [line] (x1)-- ($(x1.south)+(0,-0.25)$) -|(b1c3);


\path [line] (b3)--($(b3.south)+(0,-0.25)$) -|(b3c1);
\path [line] (b3)--($(b3.south)+(0,-0.25)$) -|(b3c2);

\path [line] (n1)--($(n1.south)+(0,-0.25)$) -|(n1c1);
\path [line] (n1)--($(n1.south)+(0,-0.25)$) -|(n1c2);

\path [line] (m1)--($(m1.south)+(0,-0.35)$) -|(m1c1);
\path [line] (m1)--($(m1.south)+(0,-0.35)$) -|(m1c2);
\path [line] (m1)--($(m1.south)+(0,-0.35)$) -|(m1c3);

\end{tikzpicture}

	\small \caption{Overview of differential privacy integration with smart grid, cloud computing, crowdsensing, data publishing networks, and healthcare applications operating over blockchain network.}
     \label{fig:tnfig01}
\end{figure*}


\section{Integrating Differential Privacy Mechanism in Various Blockchain Scenarios}\label{newsection}

Blockchain is emerging as one of the most promising technology that has the potential to raise communication, storage, and transparency in transactions to the next level. According to Tractica (a market intelligence firm), the annual revenue generated via certain enterprise blockchain applications can reach up to U.S \$ 19.9 billion at the end of year 2025~\cite{newref01}. Blockchain technology is paving its paths in multiple industrial and academic domains, such as machine learning, smart grid, cloud computing, crowdsensing, and healthcare. Certain experiments have been conducted by researchers to study the effect of integration of blockchain in these domains, and amazingly maximum of experiments outperformed their expectations from perspective of security, transparency, and data storage. With the success of these experiments, many industries have started practical implementation and also started shifting their traditional storage to blockchain based storage in order to facilitate their users with the maximum possible facilities~\cite{newref02}. \\
However, as discussed earlier, blockchain itself is a good option to enhance security and trust, but it is not pre-equipped with any privacy preservation technology. Blockchain provides pseudo anonymity via key encryption, although this pseudo anonymity is not sufficient  enough to provide complete privacy guarantee~\cite{newref03}. Therefore, there is a huge need to integrate external privacy preservation strategy before practical implementation of blockchain. In order to do so, we discussed the integration of differential privacy with blockchain, as dynamic nature and strong theoretical basis of differential privacy can protect privacy of blockchain efficiently. In this section, we discuss certain projects and researches which integrated differential privacy in their scenarios.


\begin{table*}[t!]
\begin{center}
 \centering
 \makegapedcells
 \scriptsize
 \captionsetup{labelsep=space}
 \captionsetup{justification=centering}
\caption{\textsc{\\ Overview of integration of differential privacy protection strategies in blockchain based scenarios}}
  \label{tab:sgtab01}
  {\color{black} \begin{tabular}{|P{0.9cm}|P{1.2cm}|P{0.42cm}|P{1.6cm}|P{2.1cm}|P{1.1cm}|P{1.3cm}|P{1.3cm}|P{1cm}|P{0.9cm}|P{1.6cm}|}
 \hline
\rule{0pt}{3ex}
\bfseries Major Domain & \bfseries Sub-Category & \bfseries Ref \# & \centering \bfseries Technique Name & \centering \bfseries DP Mechanism Integration & \bfseries Noise Type & \bfseries Privacy Enhancements & \bfseries General Enhancements & \bfseries Block- \newline chain \newline Type & \bfseries Consen-\newline sus & \bfseries Considered Attacks  \\
\hline



\cline{2-11}



\multirow{3}{*}{}

\rule{0pt}{3ex}
\centering \bfseries Smart Grid & \centering Energy trading & ~\cite{tnseref03} & Private energy trading in SG & A mechanism to achieve effect of DP & Structural bounded noise & \tabitem Prevented linkability w.r.t energy degree & Time Cost &  Conso- \newline rtium \scriptsize{(permissi- \newline oned)} & Self \newline designed & \tabitem Data mining \newline attacks \newline \tabitem Linking attacks \newline \tabitem Identity attacks  \\
\cline{2-11}

\rule{0pt}{3ex}
& \centering Private Data Sharing & ~\cite{tnseref04} & Fair data sharing in private smart grid & Use pagerank algorithm to deduce private reputation & Laplace & $-$ &\tabitem Customer participation \newline \tabitem Gas consumption &  Public \scriptsize{(permissi- \newline onless)} & PoA & \tabitem Double spending attack  \\
\cline{2-11}
\hline

\multirow{4}{*}{}

\rule{0pt}{3ex}\centering \bfseries Cloud Computing & \centering Cloud Federation &~\cite{tnseref05} & Differential privacy based data sharing & Autonomous smart contract based privacy budget allocation & Laplace & $-$ &\tabitem Budget consumption \newline \tabitem Workload &  Private \scriptsize{(permissi- \newline oned)} & pBFT & \tabitem Re-identification attack \\
\cline{2-11}

\rule{0pt}{3ex}
& \centering Federated Learning based Edge &~\cite{tnseref06} & Private survey feature extraction using DP and FL & DP noise is added with extracted features & Laplace & \tabitem Utility Enhancement by Increasing Test accuracy & $-$ &  Public \newline \scriptsize{(permissi- \newline onless)} & PoW & \tabitem Model poisoning attack \\
\cline{2-11}

& \centering \cont{Private Edge based IIoT} & ~\cite{techref01} & \cont{DP based Edge for IIoT} & \cont{Noise addition in energy cost set} & \cont{Laplace} & $-$ &\tabitem \cont{Time cost} \newline \tabitem \cont{Gas cost} &  \cont{Public \scriptsize{(permissi- \newline oned)}} & \cont{Not Mentioned} & \tabitem \cont{Data-Mining based attacks} \\
\cline{2-11}
\hline

\rule{0pt}{3ex}
\centering \textbf{Crowdsensing} & \centering Confidential smart contract &~\cite{tnseref07} & Aggregating \& storing private crowdwisdom & Integrating zero-knowledge proofs and DP to protect privacy & Geometric distribution $Geom(\alpha)$ & $-$ & \tabitem Computation time \newline \tabitem Gas cost & Public \scriptsize{(permissi- \newline oned)} & PoW & \tabitem Statistical attacks \newline \tabitem Side-channel attack \\

\hline

\multirow{3}{*}{}
\rule{0pt}{3ex}
\centering \textbf{Healthcare} & \centering Private healthcare 4.0 &~\cite{tnseref08} & Secure blockchain based healthcare & Discussed integration of DP & $-$ & $-$ &\tabitem Efficient maintenance \& storage & Public \newline \scriptsize{(permissi- \newline onless)} & Proof \newline of \newline Votes &$-$  \\
\cline{2-11}
& \centering \cont{Disease Diagnosis} & ~\cite{techref02} & \cont{Health Chain} & \cont{Combination of DP and Pseudo-identity} & \scriptsize{$\mu (a) \propto \newline e^{(-\beta \parallel a \parallel_2)}$} & \tabitem \cont{Test error} & $-$ &  \cont{Public \scriptsize{(permissi- \newline oned)}} & \cont{PoW} & \tabitem \cont{Identity attack}  \\
\cline{2-11}
\hline

\multirow{3}{*}{}

\rule{0pt}{3ex}
\centering \bfseries \cont{Data Publishing Networks}  & \centering Controlled data sharing \& reputation management &~\cite{tnseref09} & DP based controllable data sharing model & Control policy working over DP phenomenon & Laplace & \tabitem Reputation score & \tabitem Price balance &  Public \newline \scriptsize{(permissi- \newline onless)} & PoC &  $-$  \\
\cline{2-11}

\rule{0pt}{3ex}
& \centering Infrast- \newline ructure data protection & ~\cite{tnseref10} & Enhanced access control for infrastr- \newline ucture data via DP & DP layered protection used for aggregation & Laplace & $-$ &\tabitem Processing time &  Public \newline \scriptsize{(permissi- \newline onless)} & PoW & \tabitem Query attack \\
\cline{2-11}
\hline

 \end{tabular}}
  \end{center}
\end{table*}

 \subsection{Projects Considering Integrating of Differential Privacy with Blockchain based Smart Grid}\label{grid}
Recent advances regarding deployment and development of smart grid has opened numerous research and industrial challenges. One of such challenge is to effectively manage and perform all operations of smart grid such as communication, energy trading, renewable energy management, etc~\cite{newref08}. Researchers are actively working to overcome these challenges and are transforming smart grid to cope-up with all mentioned issues. One possible solution to effectively manage smart grid operation is its integration with blockchain technology.\\
Many possible scenarios are beings explored to integrate blockchain technology with smart grid. For example, blockchain is deployed at certain layers of smart grid to provide security to its users, such as consumption layer, generation layer, etc. Recently, a case study regarding deployment of blockchain based micro-grid in Kazakhstan is presented in researchers in~\cite{newref09}, in which they discussed the energy trading possibilities of Kazakhstan using blockchain. Literature shows that a lot of works are discussing integration of smart grid with blockchain, however plenty of works are neglecting the need of privacy preservation in this scenario. Blockchain is a publicly distributed ledger, and this raises the need of integration of privacy protection in such model. Majority of operations performed in smart grid scenarios comes under the field of real-time data analytics, therefore integrating modern noise addition mechanism of differential privacy seems to be one of the most prospective solution to overcome these challenges.\\
One such work to provide private energy trading in modern blockchain-based smart grid scenario is carried out by authors in~\cite{tnseref03}. The authors proposed their own private energy trading model by following the basic implementation details of differential privacy and compared their proposed models with existing differential privacy approaches. \cont{The proposed mechanism works over phenomenon of blockchain-based token bank to store and carry out transactions during energy trading. Similarly, the mechanism achieves effects of centralized differential privacy by preventing linkability and overcoming datamining and linking attacks along with consuming minimal computational power.} Furthermore, another work integrating central differential privacy in deregulated smart grids operating over blockchain is provided in~\cite{tnseref04}. \cont{The authors worked over enhancement of proof-of-authority (PoA) mechanism via integrating it with PageRank mechanism to formulate reputation scores.} \cont{Moreover, the authors added Laplace noise to protect users' privacy in order to encourage more participating users.} Authors claimed that their proposed strategy enhances trust by overcoming similarly, and double spending attacks of blockchain-based smart grid users.\\
The above discussion illustrates that privacy requirements should seriously be considered while integration of blockchain with smart grid, and more research efforts are required to provide smart grid users a trustable atmosphere.

\subsection{Integrating Differential Privacy with Blockchain-based Cloud Computing}\label{cloud}
The paradigm of cloud is being used by industries since long time, however researchers are enhancing this paradigm day by day and are moving towards more modern and advanced cloud computing models. One such model is the use of edge/fog computing to benefit cloud by providing quick access to important tasks~\cite{newref10}. Another model proposed by researchers is to extract features of cloud by using machine learning algorithms~\cite{newref11}. Similarly, works are being carried out to integrate blockchain with edge computing in order to provide reliable storage, control and network access along with providing the functionality of large scale network servers~\cite{newref01}. These works motivated many researchers to explore the field of blockchain-based edge and cloud computing and many researches are being carried out to enhance its efficiency and time-delay~\cite{newref12}. \\
On the other hand, some researchers also pointed out the flaw of privacy leakage in blockchain-based cloud systems and highlighted certain privacy issues in this implementation~\cite{newref13}. In order to overcome these issues, researchers are integrating privacy protection strategies with blockchain-based cloud and integrating differential privacy with blockchain-based cloud is one of a prospective solution. \cont{One such work that integrated local differential privacy with data sharing with federation based decentralized cloud is carried out in~\cite{tnseref05}.} The authors integrating the emerging concept of cloud federation with differential privacy in order to allocate autonomous privacy budget during blockchain mining. The proposed work enhanced workload during query execution and claimed that the given mechanism answers query more effectively along with protecting privacy. Private/permissioned blockchain model is used by researchers along with byzantine fault tolerant (BFT) consensus mechanism to ensure the cooperation and control by some specific authorized nodes. Furthermore, the authors claimed that the proposed mechanism successfully tackles all re-identification attacks due to its data perturbing nature. \cont{Similarly, a work that covers the domain of edge computing, centralized differential privacy, and IIoT was carried out by Gai~\textit{et al.} in~\cite{techref01}. Authors added differentially privacy Laplace noise in energy cost set to protect private values. Along with this, authors worked over enhancement of time cost and energy cost to make the algorithm less complex for small IIoT devices. Furthermore, authors claimed that the proposed  work successfully overcome all type of data-mining attacks and provided IIoT users a secure platform where they can interact without the risk of losing their private information in such attacks.}\\
\cont{Another domain of federated learning based edge computing is explored by Zhao~\textit{et al.} in~\cite{tnseref06}. The authors extracted survey features from crowd edge nodes using local differential privacy and federated learning approach along with ensuring that none of the private data of IoT users is analysed.} Furthermore, Laplace noise is added to extracted features of crowdsensing before mining the results to blockchain. The provided mechanism overcome model-poisoning attack along with enhancing test accuracy for blockchain feature extraction. From the above discussion, it can be seen that blockchain-based edge and cloud computing is not completely secure and private. Therefore, researches should be carried out to enhance privacy in such decentralized cloud scenarios.

\subsection{Integration of Differential Privacy in Crowdsensing Operating over Blockchain}\label{crowd}
The domain of crowdsensing was introduced to collect data from sensor-rich IoT devices, in order to carry out behavioural analysis and certain other similar tasks. This domain of crowdsensing strengthened its roots because of its vast advantages in the field of healthcare, environmental monitoring, and intelligent transportation~\cite{newref14}. \cont{Similarly, crowdsensing empowers the individuals by giving them right to contribute their data for various scientific experiments, in this way users can also participate in these experiments~\cite{crowd01}. Since large number of people are involved in this, the data could be noisy and cannot give clear results if collected in an uneven manner without any massive arrangement. Because of this, researchers also explored the domain of integration of crowdsensing with emerging blockchain technology, and this exploration provided to be a type of ideal exploration. Crowdsensing heavily relies over the willingness of the participants, as one will only be able to collect data if the participating individual gives permission of it. For example, a person “X” will only contribute its data when he is 100\% sure that none of his private information will get leaked. The immutable nature of blockchain ledger ensured security and tamper-resistance of data. However, on the other side, this technology also poses certain risks related to data confidentiality and privacy. The public available of decentralized ledger raises serious doubts regarding the privacy of data rich sensors and users are raising questions regarding confidentiality of their data.} \\
\cont{In order to overcome this catastrophe, researchers are working over} integrating advances privacy preserving strategies with crowdsensing in order to provide a trustworthy atmosphere to its users. One such step is the integration of differential privacy with crowdsensing, which is carried \cont{out by Duan~\textit{et al.} in~\cite{tnseref07}. Moreover, the authors worked over privacy preserving} aggregation and storage of crowd wisdom using centralized differential privacy. Furthermore, the authors also integrated zero-knowledge-proofs with differential privacy perturbation to ensure further confidentiality of data. The presented model claims to enhance computational power along with reducing gas cost spent during block mining. The authors ensured that their proposed strategy efficiently overcome statistical and side-channel attacks due to differential privacy and zero-knowledge-proofs protection. \\
After critically viewing all the discussion, it is clearly evident the differential privacy protects privacy of blockchain based crowdsensing nodes in an efficient manner. Therefore, researches need to focus over integrating modern decentralized crowdsensing scenarios with differential privacy.

\subsection{Integrating Differential Privacy in Decentralized Healthcare}\label{health}
Healthcare 4.0 is considered to be one of the core part of modern smart cities, in which every patient, doctor, and hospital will be connected with each other in order to perform certain functions such as remote health monitoring, fitness programs, and elderly care, etc~\cite{newref15}. However, because of quick urbanization, traditional healthcare devices and systems are not capable enough to meet demands and requirements of citizens. Similarly, traditional healthcare systems do not provide enough transparency and trust because they can be tampered, and data can be changed via some adversarial attacks. Therefore, trend of integration of blockchain with healthcare is increasing day by day and many hospitals and healthcare centres have started implementing blockchain based healthcare. This trend has provided numerous benefits, although it also raises some serious privacy concerns because of public availability of data. As data over blockchain is stored in a decentralized distributed ledger and every node has a copy of that ledger, therefore, some malicious node can intrigue into private data of a blockchain node.\\
Researchers are actively working to integrate privacy preservation strategies with blockchain based healthcare systems. One of such effort involves integration of differential privacy in decentralized healthcare, in which private data of patients is efficiently perturbed in order to protect their privacy. The authors in~\cite{tnseref08} proposed a secure blockchain based healthcare system operating over proof of votes consensus mechanism. \cont{The authors further added that they added noise in their data using centralized differential privacy protection to ensure user privacy.} From the above discussion it can be seen that blockchain-based healthcare required additional privacy preservation mechanism to protect users’ privacy. However, no such work that provide simulation based analysis over protecting healthcare records privacy is available in literature yet. Therefore, researches need to be carried out in this particular domain, in order to provide healthcare users a trusted and private atmosphere. \\
\cont{A detailed technical work that targets the domain of disease diagnoses has been carried out Chen~\textit{et al.} in~\cite{techref02}. Authors used the concept of machine learning and AI to learn from the hospital records, however, they kept the complete process private and secure by using centralized differential privacy and pseudo-identity mechanisms. Furthermore, the authors enhanced test errors and overcome identity attacks in a public blockchain environment. After analysing all discussion, it can be seen that the field of healthcare has been explored by few researchers, although these is plenty of room that still needs to be filled. Therefore, researches should be carried out to integrate differential privacy in blockchain based healthcare.}

\subsection{Data Publishing Networks}\label{misc}
\cont{With the advancement in modern communication and storage technologies, network data is being stored and is utilized in almost all every domain. Generally, data from all our resources such as mobile, social media, smart watches, body sensors, etc is collected and is further used to carry out statistical analysis. This collected data is said to be rich data due to its contents and is considered valuable to plenty of industries. However, on the other hand this data contains a significant amount of sensitive information that can leak privacy in it gets published without protecting it via any privacy preservation stragey~\cite{cutref02}. In order to protect this data, differential privacy can play a vital role. For example, various query protection mechanisms of differential privacy can be integrated with data publishing in order to make this data private.}\\
One such work is carried out by Dong~\textit{et al.} in the field of balanced and self-controllable data sharing in blockchain using differential privacy. The authors worked over control policy phenomenon and designed it using basics of centralized differential privacy perturbation. Authors enhanced price balance and ensured that users can have self-control over their private data, so that they can use blockchain without the fear of losing their private information. Anther similar work that integrated differential privacy with infrastructure management domain is carried out by Alnemari~\textit{et al.} in~\cite{tnseref10}. Author enhanced query processing time in decentralized blockchain network and provided a layer-wise perturbation structure for centralized differential privacy integration with infrastructure management data. After viewing the discussion, it can be concluded that there are still certain domains in blockchain systems that require considerable attention from perspective of their privacy protection.

\subsection{Summary and Lessons Learnt}
Blockchain based systems and scenarios such as machine learning, smart grid, healthcare, cloud computing, and crowdsensing  are gaining stability day by day and a lot of researches are being carried out to enhance their efficiency, throughput, privacy, and security issues. A major issue that these systems is facing is the leakage of privacy due to their transparent and publicly available nature. Therefore, researchers are moving towards addition of privacy preservation strategies with these systems. In doing so, researchers have analysed many privacy preservation mechanisms in blockchain scenarios and provided their technical and theorical pros and const. Keeping in view their implementation and protection outcomes, it can be said that differential privacy comes out to be one of the most optimal method to protect privacy of blockchain systems.\\
The discussion above provided a brief overview of integration of differential privacy in certain blockchain domains, and it also shows that researches are now being conducted to enhance differential privacy in order to fit into decentralized scenarios completely.  However, still a lot of room is left and there is a need for researchers to focus over protection and enhancement of privacy in blockchain based scenarios.\\
\cont{Apart form these, some works also highlighted the implementation of differential privacy in certain practical projects. For example, authors in~\cite{prac01} presented a detailed literature review regarding implementation of differential privacy in practical perspective. The article first inspected basic definitions of differential privacy and then moved towards state-of-the-art explanations and implementations of it in various datasets and domains. Similarly, another work named as “Differential Privacy in the Wild” provided detailed tutorial on current practices and project being carried out in the domain of differential privacy~\cite{prac02}. However, to the best of our knowledge the practical implementation of differential privacy in blockchain scenario has not been yet carried out by developers.}\\
\cont{Furthermore, some works also highlighted the use of machine learning with differential privacy in blockchain networks. For example, machine learning is being used as tool to develop promising solutions to our problems by getting deeper insights of available data in almost every field such as, bioinformatics, finance, and agriculture, and wireless communication \cite{newref04}. Similarly, privacy preserving machine learning has been explored by some researchers to provide certain useful algorithm, and these algorithms can further be integrated with multiple applications.} \\ 
\cont{For instance, Chen~\textit{et al.} in~\cite{tnsref01} proposed a differential privacy based decentralized machine learning approach which protects users’ privacy while carrying out machine learning using stochastic gradient descent (SGD). Authors named the proposed strategy as “LearningChain” and claimed that their proposed strategy provided private learning along with reducing error rate. The presented strategy works over the phenomenon of perturbing normalized local gradient information before mining it into blockchain, in this way the data is protected before making it tamper-proof, and only the desirable protected record is mined into blockchain network. Furthermore, the authors used a public blockchain and carry out consensus using proof-of-work (PoW) consensus mechanism. Moreover, to protect the system from byzantine attacks, the authors worked over~\textit{l}-nearest aggregation algorithm, that protects private data before during the collection by making it indistinguishable from its neighbours. The complete model was developed over Ethereum network and is studied using MNIST~\cite{newref06}, and Wisconsin breast cancer datasets~\cite{newref07}.}\\
\cont{Another work that discusses the integration of differential privacy in blockchain based machine learning scenario is presented by Kim~\textit{et al.} in~\cite{tnseref02}. The proposed work enhances usability and transaction latency along with protecting privacy by carrying out experiments with repeated-additive noise via differential privacy. \cont{This repeated-additive noise is used in conjunction with local gradient and is further improvised to protect blockchain user privacy.} The authors implemented a private blockchain that mines the blocks using PoW consensus mechanism. Authors claimed that they enhanced users trust in distributed machine learning by introducing efficient perturbation mechanism via differential privacy. Moreover, the authors claimed to improve users’ participation by overcoming adversarial and collusion attacks in the network. After analysing all discussion, we can conclude that \cont{differential privacy protection strategy} efficiently protects users' privacy during decentralized blockchain based machine learning scenarios.}


\begin{figure*}[t]        
\centering
\includegraphics[scale = 0.35]{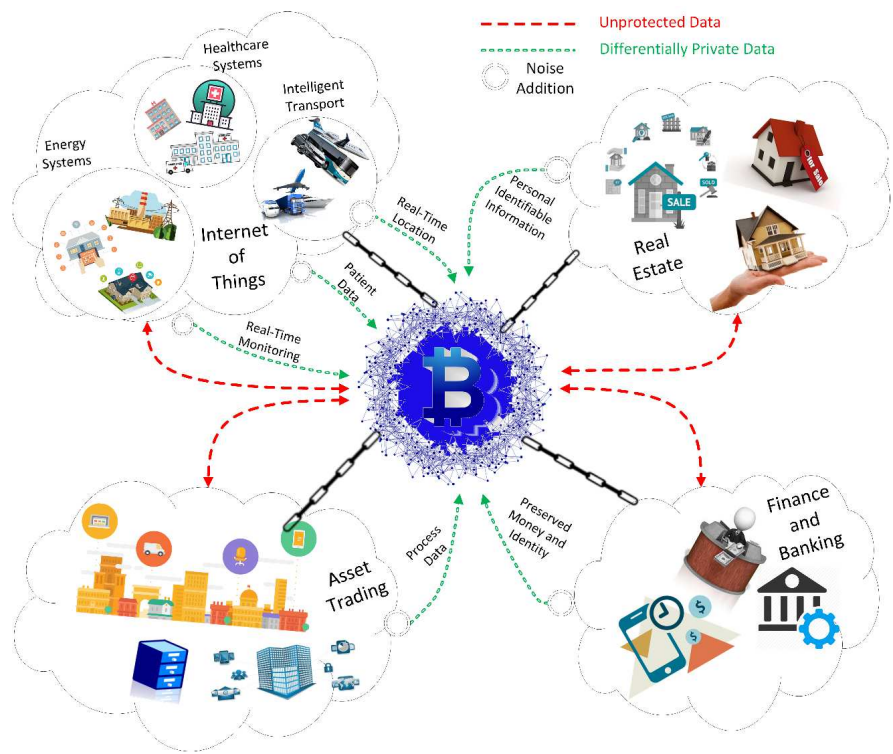}
 \caption{Application scenarios of integration of differential privacy in blockchain-based Systems (such as Internet of Things (IoT), real estate management, asset trading, and finance Trading).}      
  \label{fig:fig02}   
\end{figure*}



\begin{table*}[ht]
\begin{center}
 \centering
 \small
  \captionsetup{labelsep=space}
 \captionsetup{justification=centering}
 \caption{\textsc{\\Privacy requirements and preserved parameters of integration of differential privacy in applications of blockchain}}
  \label{tab:tstab01}
  \begin{tabular}{|P{3cm}|P{3.2cm}|P{2.5cm}|P{2.5cm}|P{4cm}|}
  	\hline
\rule{0pt}{2ex}
\bfseries Application Name & \bfseries Sub Domain & \bfseries Type of Data & \bfseries Required Privacy Level & \bfseries Parameters to Preserve  \\
\hline
\rule{0pt}{2ex}
\multirow{3}{*}{\parbox{2cm}{}}

\rule{0pt}{2ex}
& Energy Systems ~\cite{tnseref04} & Time-Series & Medium & \tabitem Real-time monitoring \newline \tabitem User identity  \\
\cline{2-5}

\rule{0pt}{2ex}
\centering \textbf{Internet of Things (IoT)} & Healthcare Systems ~\cite{tnseref08} & Time-Series \newline \& \newline Statistical & Very High & \tabitem Patients' identity \\
\cline{2-5}

\rule{0pt}{2ex}
& Intelligent Transportation Systems~\cite{lastref01} & Time-Series & Medium & \tabitem Vehicle location \newline \tabitem Driver identity \newline \tabitem Freight data  \\
\cline{2-5}
\hline

\rule{0pt}{2ex}
\centering \textbf{Real Estate} & Trading \newline (Buying \& Selling)~\cite{lastref02} & Time-Series & High & \tabitem Buyer \& seller identity \newline  \\
\hline

\rule{0pt}{2ex}
\centering \textbf{Asset Management} & Virtual or Physical Resources~\cite{lastref03} & Time-Series & Medium & \tabitem Package location \newline \tabitem Processing steps \\
\hline

\rule{0pt}{2ex}
\centering \textbf{Finance and Banking} & International Transfer~\cite{lastref04} & Time-Series & Very High & \tabitem Sender \& Receiver identity \newline \tabitem Exchanged amount \\
\hline

 \end{tabular}
  \end{center}
\end{table*}


\section{Future Applications of Differential Privacy in Blockchain}\label{blchappdp}
The blockchain has the ability to transform the stability, transparency, and security of daily networks, provided the fact that is should only be applied to the required applications, because integrating blockchain not a panacea for every problem of centralized network. Similarly, while applying blockchain to any application, its privacy protection should be taken care of. In this section, we discuss privacy protection of few applications of blockchain using differential privacy. Figure~\ref{fig:fig02} depicts the integration of differential privacy in blockchain applications to protect data privacy. Similarly, Table~\ref{tab:tstab01} gives a brief overview about the privacy requirements and parameters needed to preserve while integrating differential privacy in applications of blockchain.

\subsection{Internet of Things}
\jpdc{The notion of IoT was first introduced to address and manage devices that were connected using radio-frequency identification (RFID) wireless technology.} With the passage of time, this IoT concept shifted from RFID technology to Internet based connections. Currently, IoT devices are on an urge to take over world in every aspects of our everyday life. The scope of IoT is not limited only to a few selected domains, as it does accumulate almost every domain in which a physical device is connected to Internet or any other device using Internet protocol (IP). Few common examples of IoT systems include farming, modern energy systems (smart grid), healthcare systems, transportation systems, etc. Similarly, further advancements in IoT systems involves its integration with blockchain technology. Among IoT applications, blockchain can be integrated into applications that involve sensing, identity management, data storage, real-time data transmission, wearables, supply chain management, and other similar scenarios. Certain companies such as IBM took a step ahead and talked regarding blockchain as a technology to democratize IoT future~\cite{dpab01}. However, the data exchange between these P2P IoT networks also raises large number of privacy issues. For instance, all blockchain users are identified with the help of their public key or their hash, which can be used to track all transactions because public key is not anonymous. Therefore, protecting blockchain IoT data using efficient privacy protection strategy (such as differential privacy) is important. In this section, we discuss three major application scenarios regarding integration of differential privacy in IoT systems via blockchain technology.
\subsubsection{Energy Systems (Smart Grid)}
One of the most critical application of blockchain-based IoT system is energy sector. \jpdc{Integration of blockchain-based technology in energy systems can be applied to large number of scenarios such as smart meter usage reporting, dynamic billing, energy trading, microgrid auctions, firmware updates, individual device consumption information, grid utility data aggregation, etc.} For example, authors in~\cite{dpab02} discussed that blockchain-based decentralized energy systems will be intelligent enough to pay for consumption of energy from each device. This basically can eradicate the need of a central entity to collect and distribute bills among houses/buildings. \jpdc{Furthermore, work carried out in~\cite{jpdcref02} demonstrates the use of blockchain in energy trading via electric vehicles based smart grid. Moving towards privacy preserving blockchain, a work using permissioned blockchain along with specific signature based criterions for privacy have been published by authors in~\cite{jpdcref02}. After analysing these works, it can be seen that researchers are working over integration of privacy and blockchain with smart grid. However,} by looking upon the shared nature of blockchain-based energy systems, it can easily be said that the identities of energy devices / smart meters are not protected. Furthermore, the owner/user of particular device/meter can also be tracked by using public broadcast data. \jpdc{Therefore, any adversarial attack such as linkage or correlation attack can easily break the pseudonym based privacy guarantee of blockchain. In this way, blockchain based smart grid networks can face numerous privacy challenges, such as energy transaction linkability, auction price leakage, etc~\cite{jpdcref01}. To cope with these challenges, one of the most effective way is the integration of an efficient privacy preserving mechanism (such as differential privacy) in such networks. 
\paragraph{Case Study}
As an illustrative example, one can take the case of blockchain-based decentralized smart meter network, in which every smart meter is calculating its billing information after every 10 minutes and is reporting the transaction details publicly on ledger in order to keep the track.} From this data, any intruder can predict the identity and then can check the usage patterns of that specific house. These patterns can be used by adversary to plan any illegal activity, such as theft, etc. Smart meter cannot even stop transmitting this real-time data, because this data is used by grid utility for certain calculations, such as demand response, future load forecasting, etc. However, if we will integrate differential privacy in this scenario, the risks factor can easily be reduced to minimum. Differential privacy integrated with blockchain-based smart meter will resist meter from reporting the accurate instantaneous value of meter reading, instead it will perturb the consumption value with some calculated noise according to the requirement. So, the smart meter will be reporting a new value in which noise is added. That is why, even if any adversary gets access to this reported data, it will not be able to make a confident guess regarding the usage of particular home. Similar concept of differential privacy can also be applied over different scenarios of integration of blockchain in energy systems. However, the amount of added noise depends upon the allowed error-rate by utility/user. Hence, while integrating differential privacy with blockchain-based shared energy systems, this trade-off between accuracy and privacy needs to be addressed efficiently.

\subsubsection{Healthcare Systems}
\jpdc{Due to swift urbanization, traditional healthcare systems are not capable enough to fulfil all the required demands and necessitates of citizens. Therefore, it is becoming a need to replace these traditional systems with more technologically advanced ones. Nowadays healthcare is not just restricted to traditional hospital setup, but has moved to a completely new level in which multiple health related devices (such as smart watch, health bands, real-time ECG monitors, etc) communicate with each other to form a network also named as smart healthcare system~\cite{jpdcref04}. This smart healthcare system can contains critical data of patients that can help doctors and nurses to examine, investigate, and judge any specific medical condition even from a distant place. Since, the health records databases are very personal and critical along with, as change in just a minor attribute can risk life of some patient. Therefore, it is important to protect these systems from any type of adversaries. In order to enhance security and trust in healthcare systems, the trend of blockchain based smart healthcare systems is increasing rapidly~\cite{jpdcref05, jpdcref06}. \\
This is a very healthy integration as immutable and secure nature of blockchain will help patients and hospitals in control the  usage and sharing of their data only to specific authorities. However, the problem does not solves here, the transparent nature of blockchain can also become a viable threat to privacy of patients as the records will always remain there, and any malicious blockchain user will be able investigate and intrude into these records. Referring to all this discussion, it can be concluded that integration of a privacy preservation strategy is important before integration of blockchain in everyday healthcare domain. In order to do so, differential privacy based privacy preservation in smart decentralized healthcare can play an active role. Some works integrating  differential privacy with blockchain based smart health have been proposed in the literature. However, plenty of areas are yet to be covered in this aspect that require further privacy preservation. 
\paragraph{Case Study}
\cont{For example, in a blockchain-based medical industry,} the complete procedure from manufacturing to hospital storage is transparent and is accessible for general public.} This complete idea seems quite appealing, but it comes up with certain privacy risks for both, the manufacturing company, and the public. In this way, any unlicensed company can get to know the exact manufacturing ingredients, experimental environment, and transportation conditions of process and can produce a similar medicine with identical name to confuse the buyers. However, this condition can easily be protected using differential privacy strategies. For instance, while reporting the ingredients or the medicine number in blockchain ledger, differential privacy algorithm can protect the exact ingredient number or the device name by perturbing it with an optimal noise, so that any intruder will not be able to make accurate assumptions regarding the presence or absence of any specific ingredient in the medicine. However, the actual manufacturing details are stored in the private blockchain database, in order to backtrack in any uncertain case.

\subsubsection{Intelligent Transportation System}

Advancements in intelligent transportation system (ITS) has given birth to numerous new fields \jpdc{such as vehicle communication management, decentralized transportation system, etc,. As it is essential for modern smart vehicles to have full time Internet access in order to communicate with each other regarding their surroundings and other transportation pattern updates~\cite{jpdcref07}, therefore, researchers  working in ITSs are aiming to provide all necessary comforts to their users. In modern ITS, a smart vehicle will be a able to communicate with each other via various network interfaces (e.g. Bluetooth, WiFi, etc.), that is why decentralized and distributed nature of blockchain can be the potential technology to make this system more efficient~\cite{jpdcref08, jpdcref09 }. Furthermore, the integration of blockchain with ITSs do also solve their security risks because of its end-to-end encryption. On one hand, this fusion enhanced its security and trust, but on the other hand,} this has increased privacy risks within transportation network. For example, every vehicle will be connected to each other in a vehicular network of ITSs, and every vehicle will be exchanging different sensor’s information with each other. This communication can be made more secure by using key encryption technology of blockchain, so that nobody from outside the network will be able to decode the broadcast message. However, the users within a public blockchain can identify get information regarding other users. \jpdc{Therefore, integration of a privacy preserving strategy on top of blockchain-based ITS architecture is mandatory, and differential privacy can be the most suitable choice for it because of its dynamic nature.}
\paragraph{\jpdc{Case Study}} For instance, a car X reports its location, traffic situation, and meteorological data with time stamp. Similarly, another car Y in the blockchain network receives this information and uses it. Since, it is a public blockchain, the real identity of owner of car X can easily be revealed by using its hash value and public cryptography key. This leakage of information poses serious privacy threats to vehicle user, because its travelling routine can be observed by collecting the transmitted information. However, differential privacy integration with blockchain-based ITS can protect this information from getting leaked. Data perturbation mechanism of differential privacy can perturb the private information of vehicles’ owner in such a way that only the minimum required information is sent to broadcast after addition of noise. \jpdc{Similarly, this privacy preservation of differential privacy can also be applied to other scenarios of ITSs, such as blockchain-based railway freight and public transportation system that can be made publicly accessible along with being private in order to provide secure real-time updates to travellers.}







\subsection{Real Estate}

Dealings and transactions in real estate world needs to be transparent and opaque, but middlemen are generally required in order to do a fair deal now a day. The complete process of involving middlemen such as broker, inspectors, and notaries public is cumbersome and expensive. In order to overcome this situation, researcher community is working over implementation of blockchain-based real estate setup. As the first motto of a blockchain based system is transparency and security, so these types of systems would totally eliminate the need of middlemen for security purpose. The decentralized public ledger of blockchain will allow the sellers to advertise their properties using broadcast in the network, and similarly buyers can select their desired properties, contact sellers, make transactions, and register properties with their names, and broadcast the sold notification to the network just by using blockchain-based setup. In this way, blockchain will remove the use of expensive, and cumbersome middlemen. This system will work similar to Bitcoin, which is successfully in running from past decade. \jpdc{Till now, some works integrating blockchain with real estate has been carried out by researchers, such as MultiChain~\cite{jpdcref10}, self-managing real estate~\cite{jpdcref11}, and South African blockchain models~\cite{jpdcref12}. These blockchain-based real estate systems are pretty secure and efficient, however certain privacy related uncertainties are not tackled in them.} For example, after the successful purchase of any property, or while advertising a specific property, the identities of buyer and seller should not be publicized. It is enough in a trade that only buyer and seller know each other, without the interference of any third person. \jpdc{Making pseudonym based identities will provide a sense of insecurity for people who are trading very often or are generating good revenues. In this case, just the protection using public key cryptography is not enough, because experiments have shown that identities can be tracked using hash and public keys. To protect this process, and in order to make it more secure and private differential privacy based blockchain real estate system will be a viable solution.}
\paragraph{Case Study} \jpdc{For example, broadcasting the information after a successful purchase is important in order to protect multiple transactions for same property. \cont{But on the other hand if a buyer X is regularly purchasing and selling properties and its information} is publicly available, then it could be a potential threat for him, as he will be one of the most prominent person in the sight of adversaries. Therefore, protecting this private information before recoding it to immutable public is important. For this, differential privacy integrated with decentralized real estate can outperform other privacy preserving strategies by efficiently perturbing identity and other personally identifiable information (PII) in order to preserve privacy.} So, in differentially private blockchain based real estate systems, one can broadcast the transaction information without the risk of revealing PII to public. \\
However, one of the biggest challenge in application of differential privacy in this scenario is the identification of accurate PII parameters. As there is no predetermined rule to declare that the specific piece of information is counted in PII or it is not in PII~\cite{jpdcref13}. Therefore, the identification and declaration of PII could be done via some sort of mutual agreement inside the network in which all nodes do agree. Thus, we consider that after resolution of this PII problem, implementation of differential privacy in blockchain-based real estate trading system can be an optimal solution to preserve individual privacy.

\subsection{Asset Management}
\jpdc{Asset management can be termed as a systematic step-by-step process of operating, developing, upgrading, supplying, and disposing of physical or virtual assets in the most efficient manner (e.g., minimizing risks, and cost along with maximizing revenue)~\cite{jpdcref14}. Nowadays, trends are shifting from traditional asset management systems to digital asset management  (DAM) networks, because \cont{DAM systems provide an organized platform} to perform every required tasks.  However, because of centralized nature of DAM systems, they are not considered completely secure and trustworthy, as central authority can change or alter any information at any time without notifying others. Therefore, blockchain-based decentralized asset management systems are paving their paths in this industry and are proving to be successful systems because of their transparent and secure nature~\cite{jpdcref15}. Although, this combination is vital to many domains, but one the other hand it also raises certain privacy risks, e.g., public availability of every step from production to deployment can attract intruders which in turn can harm any specific industry. Therefore, a privacy preservation strategy such as differential privacy should be integrated with this blockchain and asset management combo.
\paragraph{Case Study}
For example, in a decentralized pharmaceutical supply chain asset management system, every step during the formation process will reported to decentralized blockchain. This reporting on one hand ensure the usage of correct ingredients, but on the other hand will create certain privacy risks. For instance, an adversary can track the exact process and can replicate the asset qualities or can copy elements of a specific medicine and use it for illegal purposes, etc. The involvement of differential privacy in this asset management scenario can reduce this privacy loss risk to minimum because the data perturbation technique of differential privacy can efficiently add up noise in required information and can protect data being leaked during broadcast.} However, during data perturbation in asset management, the added amount of noise needs to be carefully addressed. Because process step and asset information are quite critical and excess noise can destroy usefulness of data, while less noise can risk privacy leakage. So, the careful analysis about required privacy needs to be taken before implementing differential privacy in blockchain-based asset management.

\subsection{Finance and Banking}

Blockchain was first introduced to deal with cryptocurrency such as Bitcoin and Ethereum, later on researchers identified various other benefits of using blockchain in other domains as well. \jpdc{However, the benefits of using blockchain in financial transactions can never be underestimated. After analysing all discussion, researches are being carried out to develop a completely decentralized banking/financial network. For example, authors in~\cite{jpdcref16} analysed the complete procedure of blockchain-based banking transactions. Similarly, researchers in~\cite{jpdcref17} worked over improvement of monitoring and lending banking transactions via blockchain technology. Furthermore, other works have also been carried out in this domain which shows the potential of blockchain in banking and financial sector. However, blockchain bank without a top-up privacy preservation strategy is  just like an open invitation to adversaries and intruders because of its transparent nature. Therefore, integration of a satisfactory privacy preservation strategy is mandatory for successful functioning of blockchain based banking systems. In order to do so, differential privacy serves as one of the most efficient strategy because it can perturb only the required data and user have control to amount of noise they want to add. 
\paragraph{Case Study}
Let us take the case of a decentralized blockchain-banking based international money transfer. The complete transfer process} takes several currencies and banks before the receiver is able to collect money. Similarly, some services such as Western Union are fast, but they are extremely expensive. The use of blockchain in finance will eradicate the unnecessary need of middlemen and will also save a lot of time and money, as everything will happen within an open blockchain environment that ensures transparency in transactions. \jpdc{Furthermore, the transactions should be clear visible to people in the network, so that nobody will be able to perform any malicious activity. However, this transparency may also cause some privacy concerns too. For example, people in the network can get to know regarding the financial dealings of a specific person, or adversaries can target a person who is doing large transactions and may use this information for any illegal purposes. Similarly, the place to receive the payment needs to be protected, so that any thief might not be able to perform robbery if a big transaction needs to be received somewhere. Keeping in view all these points, it can be said that protecting certain amount of information during financial dealings is mandatory. One of the most efficient way to protect this information is perturbing the private information smartly  using differential privacy perturbation mechanism.} Differential privacy can efficiently perturb the desired information without risking the transparency of transaction.


\begin{figure*}[ht]        
\centering
\includegraphics[scale = 0.55]{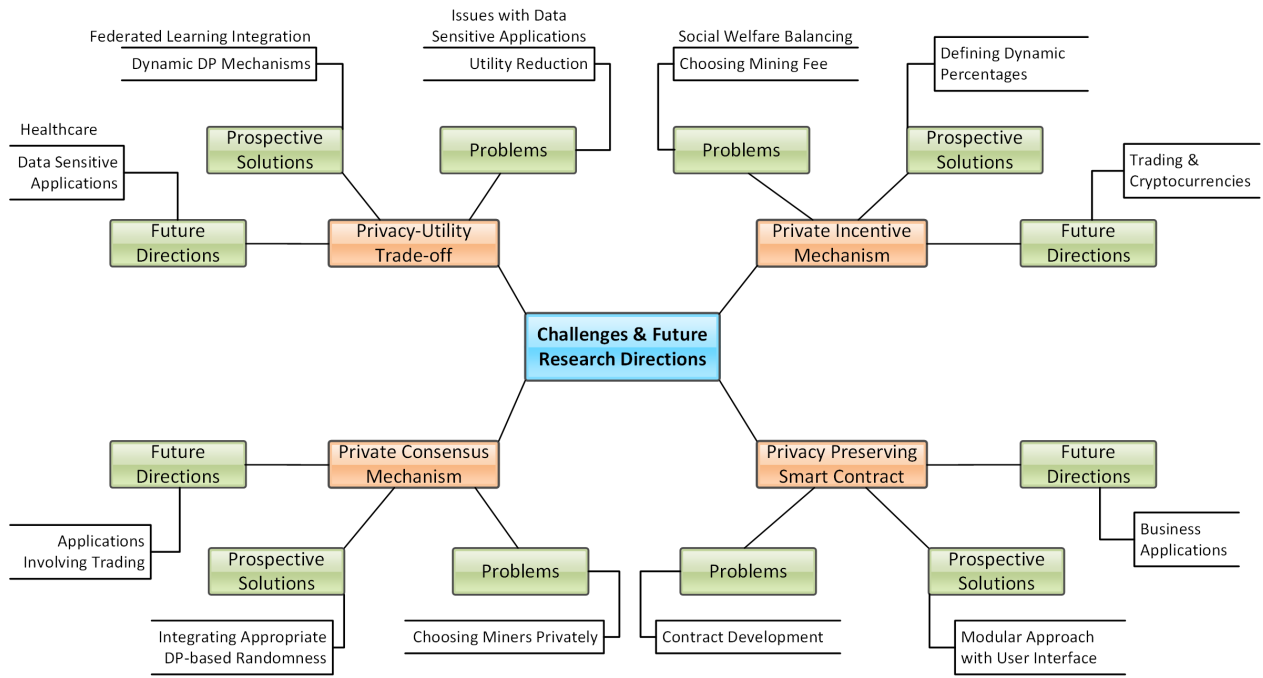}
 \caption{\textsc{\cont{Challenges, Problems, Prospective Solutions, and Future Research Directions Regarding Integration of Differential Privacy in Blockchain.}}}     
  \label{fig:challenge}   
\end{figure*}


\section{\cont{Issues, Future Directions, and Prospective Solution for Integration of Differential Privacy in Blockchain}}

\cont{In this section, we discuss issues, future directions, and prospective solutions regrading differential privacy integration in blockchain. A detailed figure regarding challenges and future research directions of integration of differential privacy in blockchain have been provided in Fig.~\ref{fig:challenge}}
\subsection{\cont{Privacy-Utility Trade-off}}
\subsubsection{\cont{Problem}}
\cont{Maintaining privacy-utility trade-off whiled integrating differential privacy is one of the most significant challenge. Noise is controlled by two parameters named as ‘epsilon’ and `sensitivity’. The careful adjustment of these parameters gives the optimal noise value according to the requirement of application. The major issue with the uneven noise is that it reduces utility of the mechanism, for example, is one adds a very high amount of noise in any query output, there is a chance that the complete query evaluation start giving false results. Similarly, during the query evaluation of decentralized blockchain networks, these parameters play the most vital role. Furthermore, certain data sensitive applications cannot even incorporate a slight level of noise, such as reporting real-time medical data. Therefore, such mechanisms need to be developed that overcome this issue in the most proficient manner.}
\subsubsection{\cont{Prospective Solutions}}
\cont{Plenty of differential privacy mechanisms have been proposed by researchers to provide minimal privacy-utility trade-off. For example, more than 50 variants of differential privacy have been highlighted by authors in~\cite{challenge01}. However, not enough literature is available from perspective of integration of differential privacy in blockchain. One prospective solution could be to combine these mechanisms and get advantage from dynamic and adaptable nature of differential privacy. Another way could be to use federated learning approach with decentralized blockchain nodes; and learn from each nodes individually. In this way, this trade-off can be adjusted dynamically, for example, if a person ‘X’ allows learning from its data with higher utility in return of some advantage from the network, then one can do it easily. However, on the other side, if a person ‘Y’ does not want to share its private information for learning, he can simply refuse it. In this way, one can also adjust its own utility-privacy trade-off according to the way that suits the best.}
\subsubsection{\cont{Future Research Directions}}
\cont{The issue of privacy-utility trade-off cannot completely be eradicated; however, it can be reduced to a maximum point by using the most appropriate mechanism of differential privacy according to the requirement. Plenty of variants of differential privacy have been proposed by researchers depending upon the need and requirement of application. For example, more than 50 variants have been highlighted by authors in~\cite{challenge01}. Similarly, in case of blockchain, researches can be carried out to propose the most optimal application orient variant of differential privacy that can effectively minimize the trade-off between utility and privacy. Once this issue gets resolved to an optimal limit, it would benefit data sensitive applications the most. For instance, healthcare personals will then be able to share their private data without risking their privacy, and query evaluators will be able to learn from this data without worrying about high utility loss.}

\subsection{\cont{Private Incentives}}
\subsubsection{\cont{Problems}}
\cont{Incentives are directly linked with money or tokens, therefore protecting privacy during incentivization can be termed as one of the most important requirement of such systems. Integrating basic noise via differential privacy in incentive layer is not a very challenging task, however, the challenge comes afterwards. The amount of noise is directly proportional to the social welfare of participants, for example, in an auction mechanism, if one adds a lot of noise in buyers bid and reduces sellers’ profit via randomization, then the participation of buyers and sellers will reduce. The accumulative profit of participants (buyers + sellers) is known as social welfare, that needs to be maintained to a specific level while carrying out trading. Similarly, choosing mining fee during the mining process can also be carried out via differentially private manner, however, choosing optimal miner fee without reducing the overall profit of the network is also a challenge that needs attention.}
\subsubsection{\cont{Prospective Solution}}
\cont{This issue of social welfare balancing and appropriate mining value selection can be encountered by defining dynamic percentages depending upon the requirement, limits, and past experiments of the network. For example, from the overall profit of the network, it can be fixed that 20-30\% will be given to buyer, 30-40\% will be given to seller, and remaining 20-30\% will be given to miner. This variation will ensure that nobody can predict the price of other with confidence. However, there still will be a need to shuffle these percentages after some time via learning from previous values and users’ behaviours. }
\subsubsection{\cont{Future Directions}}
\cont{Cryptocurrencies and banking industry is the largest domain that will get benefit after resolution of this issue. As in crypto-currencies, the mining incentives are reducing with time and there is not much interest left in mining process because of high cost and less profit. However, if miners get to know that ‘X’ company is providing more incentives than usual, then miners will always prefer the specific organization. Therefore, researches need to be carried out to develop such mechanisms that provide incentives to its users while keeping their privacy intact. So that users can easily carry out currency and asset trading without the risk of losing their private information. }

\subsection{\cont{Private Consensus}}
\subsubsection{\cont{Problem}}
\cont{Carrying out consensus in a decentralized environment was one of the functionality that made blockchain a trending paradigm. Nevertheless, basic consensus algorithms are pretty good from security perspective. However, addition of privacy is still a big challenge for researchers, and researchers are actively working over to make it more effective by using various privacy preservation mechanisms such as differential privacy, anonymization, etc. One of the biggest challenge for differential privacy researchers is to choose the miners in a private manner. For example, if one wants to choose some trading values, it can be done by randomization, however, choosing miner privately can have severe consequences in a decentralized environment in case if miner behave maliciously. Therefore, previous reputation is also required to be considered in some cases while choosing miner for mining the block. }
\subsubsection{\cont{Prospective Solution}}
\cont{Integration of appropriate level of randomness while choosing miner is the only solution for such scenario. Plenty of differential privacy mechanisms gives its users the flexibility to choose their required privacy state, and in this case, this matters a lot. However, choosing the most optimal value varies from application to application, for example, there is a possibility that a private blockchain for a cotton industry might not require very high level of privacy because all participants could be trusted, and on the other hand a public blockchain for bank could require a very high privacy. }
\subsubsection{\cont{Future Directions}}
\cont{Choosing miners in a private manner will have direct impact almost all the applications that require trading of currencies or assets, which is basically one of the largest used domain for blockchain. As privacy is considered to be the basic right of every individual and if users will be able to make a private trade even within a public environment, then they would feel more secure and comfortable to carry out large transactions without the risk of losing their data. Therefore, researches need to be carried out in development of private consensus algorithms in a way that miners will not be able to predict with confidence regarding presence or absence of a specific mining node in the mining process. This can be done by development of differentially private consensus mechanisms, in which differentially private selection of miners can be carried out via Exponential or similar mechanisms.}
\subsection{\cont{Modular Private Smart Contract}}
\subsubsection{\cont{Problem}}
\cont{Smart contracts are the basic building blocks of a decentralized blockchain network as they comprise of all set of instructions that are compulsory to run the network smoothly. However, writing and deploying private smart contract is still a challenge in majority of blockchain networks. One of the basic issue is that in order to write/modify smart contracts, one needs to have adequate knowledge of programming. Similarly, addition of differential privacy in smart contract can also be carried out after carefully analysing all network functions and then adding only required amount of perturbation every time. This require detailed knowledge of all the basic technical and theoretical concepts, which is near to impossible for a layman blockchain users. }
\subsubsection{\cont{Prospective Solution}}
\cont{In order to make integration of differential privacy easy and effective, researchers are required to develop systems that provide modular approach for its users. An important step to make blockchain more user friendly and simple has been carried out by Hyperledger, which provides various variants such as Hyperledger Fabric, Iroha, etc, which provides its users with the modular approach and they can choose their required functionalities easily. Similarly, another great step to provide modular approach for blockchain has been carried by Microsoft Azure, which provides basic drag and drop feature to its users. However, both of these technologies do not provide feature of privacy preservation via differential privacy. Therefore, researches need to be carried out that provide such modular approach to its users.}
\subsubsection{\cont{Future Directions}}
\cont{The field of privacy preserving smart contract has a lot of potential and plenty of works have been carried out by researchers in this field. Some works also targeted integration of differential privacy in smart contract from perspective of various applications. However, this field still require further exploration. For example, researches can be carried out to analyse dependencies of in smart contract, or in which specific aspect of smart contract privacy should be integrated, etc. Similarly, integration of differential privacy in application oriented smart contract is also a possible future direction, in which researchers can focus over smart contract of a specific application rather than targeting a generalized domain.}

\section{Conclusion}

Blockchain is an emerging technology and has a very gigantic future in the next five years. Along with these advancements, certain issues of blockchain needs to be addressed with time. One of the major issue of blockchain is its privacy concerns and information leakage in practical applications. In order to overcome information leakage and protect blockchain privacy, modern data perturbation technique named as differential privacy can be use used. In this paper, we present a brief discussion regarding the functionality of blockchain, and differential privacy by considering their operation phases and important parameters. Similarly, we provide in-depth discussion about integration of differential privacy in each layer of blockchain. Furthermore, we present a brief summary about the works that have been carried out regarding integration of differential privacy in decentralized blockchain technology. Finally, we concluded the article with discussion regarding challenges and future issues along with providing detailed analysis about practical implementation of differential privacy in blockchain-based everyday life applications such as Internet of Things, real estate, asset management and finances.

\bibliographystyle{IEEEtran}

\begin{thebibliography}{100}
\providecommand{\url}[1]{#1}
\csname url@samestyle\endcsname
\providecommand{\newblock}{\relax}
\providecommand{\bibinfo}[2]{#2}
\providecommand{\BIBentrySTDinterwordspacing}{\spaceskip=0pt\relax}
\providecommand{\BIBentryALTinterwordstretchfactor}{4}
\providecommand{\BIBentryALTinterwordspacing}{\spaceskip=\fontdimen2\font plus
\BIBentryALTinterwordstretchfactor\fontdimen3\font minus
  \fontdimen4\font\relax}
\providecommand{\BIBforeignlanguage}[2]{{%
\expandafter\ifx\csname l@#1\endcsname\relax
\typeout{** WARNING: IEEEtran.bst: No hyphenation pattern has been}%
\typeout{** loaded for the language `#1'. Using the pattern for}%
\typeout{** the default language instead.}%
\else
\language=\csname l@#1\endcsname
\fi
#2}}
\providecommand{\BIBdecl}{\relax}
\BIBdecl

\bibitem{blchref01}
S.~Nakamoto, ``Bitcoin: A peer-to-peer electronic cash system,''
  \emph{http://bitcoin.org/bitcoin.pdf}, 2008.

\bibitem{blchref02}
T.~T.~A. Dinh, R.~Liu, M.~Zhang, G.~Chen, B.~C. Ooi, and J.~Wang, ``Untangling
  blockchain: A data processing view of blockchain systems,'' \emph{IEEE
  Transactions on Knowledge and Data Engineering}, vol.~30, no.~7, pp.
  1366--1385, 2018.

\bibitem{blchref04}
T.~Salman, M.~Zolanvari, A.~Erbad, R.~Jain, and M.~Samaka, ``Security services
  using blockchains: A state of the art survey,'' \emph{IEEE Communications
  Surveys \& Tutorials, in Print}, 2018.

\bibitem{blchref05}
``Ethereum blockchain app platform. (2017),'' [Online]. Available:
  https://www.ethereum.org/.

\bibitem{chenref01}
L.~Qi, X.~Zhang, W.~Dou, C.~Hu, C.~Yang, and J.~Chen, ``A two-stage
  locality-sensitive hashing based approach for privacy-preserving mobile
  service recommendation in cross-platform edge environment,'' \emph{Future
  Generation Computer Systems}, vol.~88, pp. 636--643, 2018.

\bibitem{review01}
A.~Singla and E.~Bertino, ``Blockchain-based pki solutions for iot,'' in
  \emph{2018 IEEE 4th International Conference on Collaboration and Internet
  Computing (CIC)}.\hskip 1em plus 0.5em minus 0.4em\relax IEEE, 2018, pp.
  9--15.

\bibitem{review02}
T.~Salman, M.~Zolanvari, A.~Erbad, R.~Jain, and M.~Samaka, ``Security services
  using blockchains: A state of the art survey,'' \emph{IEEE Communications
  Surveys \& Tutorials}, vol.~21, no.~1, pp. 858--880, 2018.

\bibitem{review03}
J.~Weng, J.~Weng, J.~Zhang, M.~Li, Y.~Zhang, and W.~Luo, ``Deepchain: Auditable
  and privacy-preserving deep learning with blockchain-based incentive,''
  \emph{IEEE Transactions on Dependable and Secure Computing, in Print}, 2019.

\bibitem{review04}
M.~Ali, J.~Nelson, R.~Shea, and M.~J. Freedman, ``Blockstack: A global naming
  and storage system secured by blockchains,'' in \emph{2016 $\{$USENIX$\}$
  Annual Technical Conference ($\{$USENIX$\}\{$ATC$\}$ 16)}, 2016, pp.
  181--194.

\bibitem{review05}
C.~Fromknecht, D.~Velicanu, and S.~Yakoubov, ``Certcoin: A namecoin based
  decentralized authentication system 6.857 class project,'' \emph{Unpublished
  class project}, 2014.

\bibitem{blchref07}
J.~Herrera-Joancomart{\'\i} and C.~P{\'e}rez-Sol{\`a}, ``Privacy in bitcoin
  transactions: new challenges from blockchain scalability solutions,'' in
  \emph{Modeling Decisions for Artificial Intelligence}.\hskip 1em plus 0.5em
  minus 0.4em\relax Springer, 2016, pp. 26--44.

\bibitem{blchref08}
L.~Axon, ``Privacy-awareness in blockchain-based pki,'' \emph{University of
  Oxford}, 2015.

\bibitem{blchref09}
K.~Christidis and M.~Devetsikiotis, ``Blockchains and smart contracts for the
  internet of things,'' \emph{IEEE Access}, vol.~4, pp. 2292--2303, 2016.

\bibitem{dpref01}
C.~Dwork, ``Differential privacy,'' in \emph{Proceedings of the 33rd
  International Conference on Automata, Languages and Programming - Volume Part
  II}, ser. ICALP'06.\hskip 1em plus 0.5em minus 0.4em\relax Berlin,
  Heidelberg: Springer-Verlag, 2006, pp. 1--12.

\bibitem{dpref02}
T.~Wang, Z.~Zheng, M.~H. Rehmani, S.~Yao, and Z.~Huo, ``Privacy preservation in
  big data from the communication perspective—a survey,'' \emph{IEEE
  Communications Surveys \& Tutorials, in Print}, 2018.

\bibitem{jpdcref18}
M.~U. {Hassan}, M.~H. {Rehmani}, and J.~{Chen}, ``Differential privacy
  techniques for cyber physical systems: A survey,'' \emph{IEEE Communications
  Surveys \& Tutorials}, vol.~22, no.~1, pp. 746--789, 2020.

\bibitem{dpref03}
E.~ElSalamouny and S.~Gambs, ``Differential privacy models for location-based
  services,'' \emph{Transactions on Data Privacy}, vol.~9, no.~1, pp. 15--48,
  2016.

\bibitem{dpref04}
C.~Dwork, A.~Roth \emph{et~al.}, ``The algorithmic foundations of differential
  privacy,'' \emph{Foundations and Trends{\textregistered} in Theoretical
  Computer Science}, vol.~9, no. 3--4, pp. 211--407, 2014.

\bibitem{cutref03}
X.~Ren, C.-M. Yu, W.~Yu, S.~Yang, X.~Yang, J.~A. McCann, and S.~Y. Philip,
  ``Lopub: High-dimensional crowdsourced data publication with local
  differential privacy,'' \emph{IEEE Transactions on Information Forensics and
  Security}, vol.~13, no.~9, pp. 2151--2166, 2018.

\bibitem{cutref04}
J.~C. Duchi, M.~I. Jordan, and M.~J. Wainwright, ``Local privacy and
  statistical minimax rates,'' in \emph{2013 IEEE 54th Annual Symposium on
  Foundations of Computer Science}.\hskip 1em plus 0.5em minus 0.4em\relax
  IEEE, 2013, pp. 429--438.

\bibitem{cutref05}
P.~Kairouz, S.~Oh, and P.~Viswanath, ``Extremal mechanisms for local
  differential privacy,'' in \emph{Advances in neural information processing
  systems}, 2014, pp. 2879--2887.

\bibitem{survey01}
M.~Conti, E.~S. Kumar, C.~Lal, and S.~Ruj, ``A survey on security and privacy
  issues of bitcoin,'' \emph{IEEE Communications Surveys \& Tutorials},
  vol.~20, no.~4, pp. 3416--3452, 2018.

\bibitem{survey02}
M.~C.~K. Khalilov and A.~Levi, ``A survey on anonymity and privacy in
  bitcoin-like digital cash systems,'' \emph{IEEE Communications Surveys \&
  Tutorials}, vol.~20, no.~3, pp. 2543--2585, 2018.

\bibitem{survey03}
J.~B. Bernabe, J.~L. Canovas, J.~L. Hernandez-Ramos, R.~T. Moreno, and
  A.~Skarmeta, ``Privacy-preserving solutions for blockchain: review and
  challenges,'' \emph{IEEE Access}, vol.~7, pp. 164\,908--164\,940, 2019.

\bibitem{newref02}
M.~S. Ali, M.~Vecchio, M.~Pincheira, K.~Dolui, F.~Antonelli, and M.~H. Rehmani,
  ``Applications of blockchains in the internet of things: A comprehensive
  survey,'' \emph{IEEE Communications Surveys \& Tutorials}, vol.~21, no.~2,
  pp. 1676--1717, 2018.

\bibitem{survey06}
Q.~Feng, D.~He, S.~Zeadally, M.~K. Khan, and N.~Kumar, ``A survey on privacy
  protection in blockchain system,'' \emph{Journal of Network and Computer
  Applications}, vol. 126, pp. 45--58, 2019.

\bibitem{newref03}
M.~U. Hassan, M.~H. Rehmani, and J.~Chen, ``Privacy preservation in blockchain
  based iot systems: Integration issues, prospects, challenges, and future
  research directions,'' \emph{Future Generation Computer Systems}, vol.~97,
  pp. 512--529, 2019.

\bibitem{survey08}
Y.~Cui, B.~Pan, and Y.~Sun, ``A survey of privacy-preserving techniques for
  blockchain,'' in \emph{International Conference on Artificial Intelligence
  and Security}.\hskip 1em plus 0.5em minus 0.4em\relax Springer, 2019, pp.
  225--234.

\bibitem{survey09}
M.~{Saad}, J.~{Spaulding}, L.~{Njilla}, C.~{Kamhoua}, S.~{Shetty}, D.~H.
  {Nyang}, and D.~{Mohaisen}, ``Exploring the attack surface of blockchain: A
  comprehensive survey,'' \emph{IEEE Communications Surveys Tutorials}, pp.
  1--1, 2020.

\bibitem{survey10}
J.~Sengupta, S.~Ruj, and S.~D. Bit, ``A comprehensive survey on attacks,
  security issues and blockchain solutions for iot and iiot,'' \emph{Journal of
  Network and Computer Applications}, p. 102481, 2019.

\bibitem{trade01}
T.~Wang, Z.~Zheng, M.~H. Rehmani, S.~Yao, and Z.~Huo, ``Privacy preservation in
  big data from the communication perspective—a survey,'' \emph{IEEE
  Communications Surveys \& Tutorials}, vol.~21, no.~1, pp. 753--778, 2018.

\bibitem{trade02}
M.~U. Hassan, M.~H. Rehmani, R.~Kotagiri, J.~Zhang, and J.~Chen, ``Differential
  privacy for renewable energy resources based smart metering,'' \emph{Journal
  of Parallel and Distributed Computing}, vol. 131, pp. 69--80, 2019.

\bibitem{cutref06}
C.~Dwork, A.~Roth \emph{et~al.}, ``The algorithmic foundations of differential
  privacy,'' \emph{Foundations and Trends{\textregistered} in Theoretical
  Computer Science}, vol.~9, no. 3--4, pp. 211--407, 2014.

\bibitem{cutref01}
J.~Lee and C.~Clifton, ``How much is enough? choosing $\varepsilon$ for
  differential privacy,'' in \emph{International Conference on Information
  Security}.\hskip 1em plus 0.5em minus 0.4em\relax Springer, 2011, pp.
  325--340.

\bibitem{chenref02}
P.~Wang, J.~Huang, Z.~Cui, L.~Xie, and J.~Chen, ``A gaussian error correction
  multi-objective positioning model with nsga-ii,'' \emph{Concurrency and
  Computation: Practice and Experience}, vol.~32, no.~5, p. e5464, 2020.

\bibitem{chenref03}
X.~Cai, Y.~Niu, S.~Geng, J.~Zhang, Z.~Cui, J.~Li, and J.~Chen, ``An
  under-sampled software defect prediction method based on hybrid
  multi-objective cuckoo search,'' \emph{Concurrency and Computation: Practice
  and Experience}, vol.~32, no.~5, p. e5478, 2020.

\bibitem{dpref05}
Y.-A. De~Montjoye, L.~Radaelli, V.~K. Singh \emph{et~al.}, ``Unique in the
  shopping mall: On the reidentifiability of credit card metadata,''
  \emph{Science}, vol. 347, no. 6221, pp. 536--539, 2015.

\bibitem{dpref06}
G.~Eibl and D.~Engel, ``Differential privacy for real smart metering data,''
  \emph{Computer Science-Research and Development}, vol.~32, no. 1-2, pp.
  173--182, 2017.

\bibitem{platref01}
M.~Belotti, N.~Bo{\v{z}}i{\'c}, G.~Pujolle, and S.~Secci, ``A vademecum on
  blockchain technologies: When, which, and how,'' \emph{IEEE Communications
  Surveys \& Tutorials}, vol.~21, no.~4, pp. 3796--3838, 2019.

\bibitem{platref02}
G.~Wood \emph{et~al.}, ``Ethereum: A secure decentralised generalised
  transaction ledger,'' \emph{Ethereum project yellow paper}, vol. 151, no.
  2014, pp. 1--32, 2014.

\bibitem{platref03}
H.~Chen, M.~Pendleton, L.~Njilla, and S.~Xu, ``A survey on ethereum systems
  security: Vulnerabilities, attacks and defenses,'' \emph{arXiv preprint
  arXiv:1908.04507}, 2019.

\bibitem{platref04}
C.~Cachin \emph{et~al.}, ``Architecture of the hyperledger blockchain fabric,''
  in \emph{Workshop on distributed cryptocurrencies and consensus ledgers},
  vol. 310, 2016, p.~4.

\bibitem{platref05}
M.~Vukoli{\'c}, ``Rethinking permissioned blockchains,'' in \emph{Proceedings
  of the ACM Workshop on Blockchain, Cryptocurrencies and Contracts}, 2017, pp.
  3--7.

\bibitem{platref06}
M.~Divya and N.~B. Biradar, ``Iota-next generation block chain,''
  \emph{International Journal Of Engineering And Computer Science}, vol.~7,
  no.~04, pp. 23\,823--23\,826, 2018.

\bibitem{platref07}
``{Trinity Attack Incident: Summary and next steps [Online] Available:
  https://blog.iota.org/trinity-attack-incident-part-1-summary-and-next-steps-8c7ccc4d81e8},''
  2020.

\bibitem{layer01}
R.~Yang, F.~R. Yu, P.~Si, Z.~Yang, and Y.~Zhang, ``Integrated blockchain and
  edge computing systems: A survey, some research issues and challenges,''
  \emph{IEEE Communications Surveys \& Tutorials}, vol.~21, no.~2, pp.
  1508--1532, 2019.

\bibitem{layer02}
J.~Wu and N.~K. Tran, ``Application of blockchain technology in sustainable
  energy systems: An overview,'' \emph{Sustainability}, vol.~10, no.~9, p.
  3067, 2018.

\bibitem{layer03}
J.~Xie, H.~Tang, T.~Huang, F.~R. Yu, R.~Xie, J.~Liu, and Y.~Liu, ``A survey of
  blockchain technology applied to smart cities: Research issues and
  challenges,'' \emph{IEEE Communications Surveys \& Tutorials}, vol.~21,
  no.~3, pp. 2794--2830, 2019.

\bibitem{layer04}
M.~Wu, K.~Wang, X.~Cai, S.~Guo, M.~Guo, and C.~Rong, ``A comprehensive survey
  of blockchain: From theory to iot applications and beyond,'' \emph{IEEE
  Internet of Things Journal}, vol.~6, no.~5, pp. 8114--8154, 2019.

\bibitem{layer05}
X.~Xu, I.~Weber, M.~Staples, L.~Zhu, J.~Bosch, L.~Bass, C.~Pautasso, and
  P.~Rimba, ``A taxonomy of blockchain-based systems for architecture design,''
  in \emph{IEEE International Conference on Software Architecture (ICSA)},
  2017, pp. 243--252.

\bibitem{layer06}
K.~S.~S. Wai, E.~C. Htoon, and N.~N.~M. Thein, ``Storage structure of student
  record based on hyperledger fabric blockchain,'' in \emph{International
  Conference on Advanced Information Technologies (ICAIT)}.\hskip 1em plus
  0.5em minus 0.4em\relax IEEE, 2019, pp. 108--113.

\bibitem{layer07}
F.~Wang, Y.~Chen, R.~Wang, A.~O. Francis, B.~Emmanuel, W.~Zheng, and J.~Chen,
  ``An experimental investigation into the hash functions used in
  blockchains,'' \emph{IEEE Transactions on Engineering Management}, 2019.

\bibitem{layer08}
{Apple, D}, ``Learning with privacy at scale,'' \emph{Apple Machine Learning
  Journal}, vol.~1, no.~8, 2017.

\bibitem{layer09}
F.~R. Yu and Y.~He, ``A service-oriented blockchain system with
  virtualization,'' \emph{Trans. Blockchain Technology and Applications},
  vol.~1, no.~1, pp. 1--10, 2019.

\bibitem{layer10}
J.~Kan, L.~Zou, B.~Liu, and X.~Huang, ``Boost blockchain broadcast propagation
  with tree routing,'' in \emph{International Conference on Smart
  Blockchain}.\hskip 1em plus 0.5em minus 0.4em\relax Springer, 2018, pp.
  77--85.

\bibitem{layer11}
T.~Jin, X.~Zhang, Y.~Liu, and K.~Lei, ``Blockndn: A bitcoin blockchain
  decentralized system over named data networking,'' in \emph{Ninth
  International Conference on Ubiquitous and Future Networks (ICUFN)}.\hskip
  1em plus 0.5em minus 0.4em\relax IEEE, 2017, pp. 75--80.

\bibitem{layer12}
T.~Song, R.~Li, B.~Mei, J.~Yu, X.~Xing, and X.~Cheng, ``A privacy preserving
  communication protocol for iot applications in smart homes,'' \emph{IEEE
  Internet of Things Journal}, vol.~4, no.~6, pp. 1844--1852, 2017.

\bibitem{layer13}
P.~Danzi, A.~E. Kal{\o}r, {\v{C}}.~Stefanovi{\'c}, and P.~Popovski, ``Delay and
  communication tradeoffs for blockchain systems with lightweight iot
  clients,'' \emph{IEEE Internet of Things Journal}, vol.~6, no.~2, pp.
  2354--2365, 2019.

\bibitem{layer14}
G.~{Xu}, Y.~{Liu}, and P.~W. {Khan}, ``Improvement of the dpos consensus
  mechanism in blockchain based on vague sets,'' \emph{IEEE Transactions on
  Industrial Informatics}, vol.~16, no.~6, pp. 4252--4259, 2020.

\bibitem{layer15}
W.~Wang, D.~T. Hoang, P.~Hu, Z.~Xiong, D.~Niyato, P.~Wang, Y.~Wen, and D.~I.
  Kim, ``A survey on consensus mechanisms and mining strategy management in
  blockchain networks,'' \emph{IEEE Access}, vol.~7, pp. 22\,328--22\,370,
  2019.

\bibitem{layer16}
S.~Nakamoto \emph{et~al.}, ``A peer-to-peer electronic cash system,''
  \emph{Bitcoin.--URL: https://bitcoin. org/bitcoin. pdf}, 2008.

\bibitem{layer18}
V.~Buterin \emph{et~al.}, ``Ethereum: A next-generation smart contract and
  decentralized application platform,'' \emph{URL https://github.
  com/ethereum/wiki/wiki/\% 5BEnglish\% 5D-White-Paper}, 2014.

\bibitem{layer17}
S.~King and S.~Nadal, ``Ppcoin: Peer-to-peer crypto-currency with
  proof-of-stake,'' \emph{self-published paper, August}, vol.~19, 2012.

\bibitem{layer19}
M.~Castro and B.~Liskov, ``Practical byzantine fault tolerance and proactive
  recovery,'' \emph{ACM Transactions on Computer Systems (TOCS)}, vol.~20,
  no.~4, pp. 398--461, 2002.

\bibitem{layer20}
M.~Castro, B.~Liskov \emph{et~al.}, ``Practical byzantine fault tolerance,'' in
  \emph{OSDI}, vol.~99, no. 1999, 1999, pp. 173--186.

\bibitem{layer21}
Z.~Liu, N.~C. Luong, W.~Wang, D.~Niyato, P.~Wang, Y.-C. Liang, and D.~I. Kim,
  ``A survey on blockchain: A game theoretical perspective,'' \emph{IEEE
  Access}, vol.~7, pp. 47\,615--47\,643, 2019.

\bibitem{layer22}
E.~Androulaki, A.~Barger, V.~Bortnikov, C.~Cachin, K.~Christidis, A.~De~Caro,
  D.~Enyeart, C.~Ferris, G.~Laventman, Y.~Manevich \emph{et~al.}, ``Hyperledger
  fabric: a distributed operating system for permissioned blockchains,'' in
  \emph{Proceedings of the Thirteenth EuroSys Conference}, 2018, pp. 1--15.

\bibitem{jpdcref01}
M.~U. {Hassan}, M.~H. {Rehmani}, and J.~{Chen}, ``Deal: Differentially private
  auction for blockchain-based microgrids energy trading,'' \emph{IEEE
  Transactions on Services Computing}, vol.~13, no.~2, pp. 263--275, 2020.

\bibitem{layer23}
A.~Kosba, A.~Miller, E.~Shi, Z.~Wen, and C.~Papamanthou, ``Hawk: The blockchain
  model of cryptography and privacy-preserving smart contracts,'' in \emph{2016
  IEEE symposium on security and privacy (SP)}.\hskip 1em plus 0.5em minus
  0.4em\relax IEEE, 2016, pp. 839--858.

\bibitem{meta01}
T.~Faisal, N.~Courtois, and A.~Serguieva, ``The evolution of embedding metadata
  in blockchain transactions,'' in \emph{IEEE International Joint Conference on
  Neural Networks (IJCNN)}, 2018, pp. 1--9.

\bibitem{layer24}
R.~Yuan, Y.-B. Xia, H.-B. Chen, B.-Y. Zang, and J.~Xie, ``Shadoweth: Private
  smart contract on public blockchain,'' \emph{Journal of Computer Science and
  Technology}, vol.~33, no.~3, pp. 542--556, 2018.

\bibitem{layer25}
H.~Kalodner, S.~Goldfeder, X.~Chen, S.~M. Weinberg, and E.~W. Felten,
  ``Arbitrum: Scalable, private smart contracts,'' in \emph{27th $\{$USENIX$\}$
  Security Symposium ($\{$USENIX$\}$ Security 18)}, 2018, pp. 1353--1370.

\bibitem{layer26}
D.~C. S{\'a}nchez, ``Raziel: Private and verifiable smart contracts on
  blockchains,'' \emph{arXiv preprint arXiv:1807.09484}, 2018.

\bibitem{layer27}
G.~Hurlburt, ``Might the blockchain outlive bitcoin?'' \emph{It Professional},
  vol.~18, no.~2, pp. 12--16, 2016.

\bibitem{layer28}
Y.~Liu, F.~R. Yu, X.~Li, H.~Ji, and V.~C. Leung, ``Blockchain and machine
  learning for communications and networking systems,'' \emph{IEEE
  Communications Surveys \& Tutorials, in Print}, 2020.

\bibitem{layer29}
M.~I. Mehar, C.~L. Shier, A.~Giambattista, E.~Gong, G.~Fletcher, R.~Sanayhie,
  H.~M. Kim, and M.~Laskowski, ``Understanding a revolutionary and flawed grand
  experiment in blockchain: the dao attack,'' \emph{Journal of Cases on
  Information Technology (JCIT)}, vol.~21, no.~1, pp. 19--32, 2019.

\bibitem{tnseref07}
H.~Duan, Y.~Zheng, Y.~Du, A.~Zhou, C.~Wang, and M.~H. Au, ``Aggregating crowd
  wisdom via blockchain: A private, correct, and robust realization,'' in
  \emph{IEEE International Conference on Pervasive Computing and Communications
  (PerCom)}, 2019, pp. 43--52.

\bibitem{tnseref08}
J.~Vora, A.~Nayyar, S.~Tanwar, S.~Tyagi, N.~Kumar, M.~S. Obaidat, and J.~J.
  Rodrigues, ``Bheem: A blockchain-based framework for securing electronic
  health records,'' in \emph{IEEE Globecom Workshops (GC Wkshps)}, 2018, pp.
  1--6.

\bibitem{techref02}
X.~Chen, X.~Wang, and K.~Yang, ``Asynchronous blockchain-based
  privacy-preserving training framework for disease diagnosis,'' in \emph{IEEE
  International Conference on Big Data (Big Data)}, 2019, pp. 5469--5473.

\bibitem{tnseref03}
K.~Gai, Y.~Wu, L.~Zhu, M.~Qiu, and M.~Shen, ``Privacy-preserving energy trading
  using consortium blockchain in smart grid,'' \emph{IEEE Transactions on
  Industrial Informatics, in Print}, 2019.

\bibitem{tnseref04}
O.~Samuel, M.~A. Nadeem~Javaid, Z.~Ahmed, M.~Imran, and M.~Guizani, ``A
  blockchain model for fair data sharing in deregulated smart grids,'' in
  \emph{IEEE Global Communications Conference: Communication \& Information
  Systems Security, USA}, 2019, pp. 1--7.

\bibitem{tnseref05}
M.~Yang, A.~Margheri, R.~Hu, and V.~Sassone, ``Differentially private data
  sharing in a cloud federation with blockchain,'' \emph{IEEE Cloud Computing},
  vol.~5, no.~6, pp. 69--79, 2018.

\bibitem{tnseref06}
Y.~Zhao, J.~Zhao, L.~Jiang, R.~Tan, and D.~Niyato, ``Mobile edge computing,
  blockchain and reputation-based crowdsourcing iot federated learning: A
  secure, decentralized and privacy-preserving system,'' \emph{arXiv preprint
  arXiv:1906.10893}, 2019.

\bibitem{techref01}
K.~Gai, Y.~Wu, L.~Zhu, Z.~Zhang, and M.~Qiu, ``Differential privacy-based
  blockchain for industrial internet of things,'' \emph{IEEE Transactions on
  Industrial Informatics, in Print}, 2019.

\bibitem{tnseref09}
X.~Dong, B.~Guo, Y.~Shen, X.~Duan, Y.~Shen, and H.~Zhang, ``A self-controllable
  and balanced data sharing model,'' \emph{IEEE Access}, vol.~7, pp.
  103\,275--103\,290, 2019.

\bibitem{tnseref10}
A.~Alnemari, S.~Arodi, V.~R. Sosa, S.~Pandey, C.~Romanowski, R.~Raj, and
  S.~Mishra, ``Protecting infrastructure data via enhanced access control,
  blockchain and differential privacy,'' in \emph{International Conference on
  Critical Infrastructure Protection}.\hskip 1em plus 0.5em minus 0.4em\relax
  Springer, 2018, pp. 113--125.

\bibitem{newref01}
R.~Yang, F.~R. Yu, P.~Si, Z.~Yang, and Y.~Zhang, ``Integrated blockchain and
  edge computing systems: A survey, some research issues and challenges,''
  \emph{IEEE Communications Surveys \& Tutorials}, vol.~21, no.~2, pp.
  1508--1532, 2019.

\bibitem{newref08}
M.~H. Rehmani, M.~Reisslein, A.~Rachedi, M.~Erol-Kantarci, and M.~Radenkovic,
  ``Integrating renewable energy resources into the smart grid: Recent
  developments in information and communication technologies,'' \emph{IEEE
  Transactions on Industrial Informatics}, vol.~14, no.~7, pp. 2814--2825,
  2018.

\bibitem{newref09}
D.~Orazgaliyev, Y.~Lukpanov, I.~A. Ukaegbu, and H.~S.~K. Nunna, ``Towards the
  application of blockchain technology for smart grids in kazakhstan,'' in
  \emph{21st International Conference on Advanced Communication Technology
  (ICACT)}.\hskip 1em plus 0.5em minus 0.4em\relax IEEE, 2019, pp. 273--278.

\bibitem{newref10}
J.~Moura and D.~Hutchison, ``Game theory for multi-access edge computing:
  Survey, use cases, and future trends,'' \emph{IEEE Communications Surveys \&
  Tutorials}, vol.~21, no.~1, pp. 260--288, 2018.

\bibitem{newref11}
T.~K. {Rodrigues}, K.~{Suto}, H.~{Nishiyama}, J.~{Liu}, and N.~{Kato},
  ``Machine learning meets computation and communication control in evolving
  edge and cloud: Challenges and future perspective,'' \emph{IEEE
  Communications Surveys Tutorials, in Print}, pp. 1--1, 2019.

\bibitem{newref12}
S.~Wang, X.~Wang, and Y.~Zhang, ``A secure cloud storage framework with access
  control based on blockchain,'' \emph{IEEE Access}, vol.~7, pp.
  112\,713--112\,725, 2019.

\bibitem{newref13}
S.~Pavithra, S.~Ramya, and S.~Prathibha, ``A survey on cloud security issues
  and blockchain,'' in \emph{3rd International Conference on Computing and
  Communications Technologies (ICCCT)}.\hskip 1em plus 0.5em minus 0.4em\relax
  IEEE, 2019, pp. 136--140.

\bibitem{newref14}
X.~Zhang, Z.~Yang, W.~Sun, Y.~Liu, S.~Tang, K.~Xing, and X.~Mao, ``Incentives
  for mobile crowd sensing: A survey,'' \emph{IEEE Communications Surveys \&
  Tutorials}, vol.~18, no.~1, pp. 54--67, 2015.

\bibitem{crowd01}
Y.~Liu, L.~Kong, and G.~Chen, ``Data-oriented mobile crowdsensing: A
  comprehensive survey,'' \emph{IEEE Communications Surveys \& Tutorials},
  vol.~21, no.~3, pp. 2849--2885, 2019.

\bibitem{newref15}
J.~Xie, H.~Tang, T.~Huang, F.~R. Yu, R.~Xie, J.~Liu, and Y.~Liu, ``A survey of
  blockchain technology applied to smart cities: Research issues and
  challenges,'' \emph{IEEE Communications Surveys \& Tutorials, in Print},
  2019.

\bibitem{cutref02}
J.~H. Abawajy, M.~I.~H. Ninggal, and T.~Herawan, ``Privacy preserving social
  network data publication,'' \emph{IEEE communications surveys \& tutorials},
  vol.~18, no.~3, pp. 1974--1997, 2016.

\bibitem{prac01}
N.~Li, M.~Lyu, D.~Su, and W.~Yang, ``Differential privacy: From theory to
  practice,'' \emph{Synthesis Lectures on Information Security, Privacy, \&
  Trust}, vol.~8, no.~4, pp. 1--138, 2016.

\bibitem{prac02}
A.~Machanavajjhala, X.~He, and M.~Hay, ``Differential privacy in the wild: A
  tutorial on current practices \& open challenges,'' in \emph{Proceedings of
  the ACM International Conference on Management of Data}, 2017, pp.
  1727--1730.

\bibitem{newref04}
C.~Luo, J.~Ji, Q.~Wang, X.~Chen, and P.~Li, ``Channel state information
  prediction for 5g wireless communications: A deep learning approach,''
  \emph{IEEE Transactions on Network Science and Engineering, in Print}, 2018.

\bibitem{tnsref01}
X.~Chen, J.~Ji, C.~Luo, W.~Liao, and P.~Li, ``When machine learning meets
  blockchain: A decentralized, privacy-preserving and secure design,'' in
  \emph{IEEE International Conference on Big Data (Big Data)}, 2018, pp.
  1178--1187.

\bibitem{newref06}
L.~Deng, ``The mnist database of handwritten digit images for machine learning
  research [best of the web],'' \emph{IEEE Signal Processing Magazine},
  vol.~29, no.~6, pp. 141--142, 2012.

\bibitem{newref07}
A.~Asuncion and D.~Newman, ``Uci machine learning repository,'' 2007.

\bibitem{tnseref02}
H.~{Kim}, S.~{Kim}, J.~Y. {Hwang}, and C.~{Seo}, ``Efficient privacy-preserving
  machine learning for blockchain network,'' \emph{IEEE Access}, pp. 1--1,
  2019.

\bibitem{lastref01}
T.~{Jiang}, H.~{Fang}, and H.~{Wang}, ``Blockchain-based internet of vehicles:
  Distributed network architecture and performance analysis,'' \emph{IEEE
  Internet of Things Journal}, vol.~6, no.~3, pp. 4640--4649, June 2019.

\bibitem{lastref02}
A.~Spielman, ``Blockchain: digitally rebuilding the real estate industry,''
  Ph.D. dissertation, Massachusetts Institute of Technology, 2016.

\bibitem{lastref03}
Y.~Zhu, Y.~Qin, Z.~Zhou, X.~Song, G.~Liu, and W.~C.-C. Chu, ``Digital asset
  management with distributed permission over blockchain and attribute-based
  access control,'' in \emph{International Conference on Services Computing
  (SCC)}.\hskip 1em plus 0.5em minus 0.4em\relax IEEE, 2018, pp. 193--200.

\bibitem{lastref04}
A.~A. Alidin, A.~A.~A. Ali-Wosabi, and Z.~Yusoff, ``Overview of blockchain
  implementation on islamic finance: Saadiqin experience,'' in \emph{Cyber
  Resilience Conference (CRC)}.\hskip 1em plus 0.5em minus 0.4em\relax IEEE,
  2018, pp. 1--2.

\bibitem{dpab01}
T.~M. Fern{\'a}ndez-Caram{\'e}s and P.~Fraga-Lamas, ``A review on the use of
  blockchain for the internet of things,'' \emph{IEEE Access}, 2018.

\bibitem{dpab02}
T.~Lundqvist, A.~de~Blanche, and H.~R.~H. Andersson, ``Thing-to-thing
  electricity micro payments using blockchain technology,'' in \emph{IEEE
  Global Internet of Things Summit (GIoTS)}, 2017, pp. 1--6.

\bibitem{jpdcref02}
J.~{Kang}, R.~{Yu}, X.~{Huang}, S.~{Maharjan}, Y.~{Zhang}, and E.~{Hossain},
  ``Enabling localized peer-to-peer electricity trading among plug-in hybrid
  electric vehicles using consortium blockchains,'' \emph{IEEE Transactions on
  Industrial Informatics}, vol.~13, no.~6, pp. 3154--3164, Dec 2017.

\bibitem{jpdcref04}
J.~Xie, H.~Tang, T.~Huang, F.~R. Yu, R.~Xie, J.~Liu, and Y.~Liu, ``A survey of
  blockchain technology applied to smart cities: Research issues and
  challenges,'' \emph{IEEE Communications Surveys \& Tutorials, in Print},
  2019.

\bibitem{jpdcref05}
T.-T. Kuo, H.-E. Kim, and L.~Ohno-Machado, ``Blockchain distributed ledger
  technologies for biomedical and health care applications,'' \emph{Journal of
  the American Medical Informatics Association}, vol.~24, no.~6, pp.
  1211--1220, 2017.

\bibitem{jpdcref06}
M.~Mettler, ``Blockchain technology in healthcare: The revolution starts
  here,'' in \emph{IEEE 18th International Conference on e-Health Networking,
  Applications and Services (Healthcom)}, 2016, pp. 1--3.

\bibitem{jpdcref07}
J.~Zhang, F.-Y. Wang, K.~Wang, W.-H. Lin, X.~Xu, and C.~Chen, ``Data-driven
  intelligent transportation systems: A survey,'' \emph{IEEE Transactions on
  Intelligent Transportation Systems}, vol.~12, no.~4, pp. 1624--1639, 2011.

\bibitem{jpdcref08}
Y.~Yuan and F.-Y. Wang, ``Towards blockchain-based intelligent transportation
  systems,'' in \emph{IEEE 19th International Conference on Intelligent
  Transportation Systems (ITSC)}, 2016, pp. 2663--2668.

\bibitem{jpdcref09}
W.~Hu, Y.~Hu, W.~Yao, and H.~Li, ``A blockchain-based byzantine consensus
  algorithm for information authentication of the internet of vehicles,''
  \emph{IEEE Access}, vol.~7, pp. 139\,703--139\,711, 2019.

\bibitem{jpdcref10}
M.~Avantaggiato and P.~Gallo, ``Challenges and opportunities using multichain
  for real estate,'' in \emph{IEEE International Black Sea Conference on
  Communications and Networking (BlackSeaCom)}, 2019, pp. 1--5.

\bibitem{jpdcref11}
N.~Shedroff, ``Self-managing real estate,'' \emph{Computer}, no.~1, pp.
  104--104, 2018.

\bibitem{jpdcref12}
J.~L. Tilbury, E.~de~la Rey, and K.~van~der Schyff, ``Business process models
  of blockchain and south african real estate transactions,'' in \emph{IEEE
  International Conference on Advances in Big Data, Computing and Data
  Communication Systems (icABCD)}, 2019, pp. 1--7.

\bibitem{jpdcref13}
A.~Narayanan and V.~Shmatikov, ``Myths and fallacies of personally identifiable
  information,'' \emph{Communications of the ACM}, vol.~53, no.~6, pp. 24--26,
  2010.

\bibitem{jpdcref14}
E.~Nielsen, ``Impact on financial performance by physical asset management,''
  \emph{-}, 2015.

\bibitem{jpdcref15}
Y.~Zhu, Y.~Qin, Z.~Zhou, X.~Song, G.~Liu, and W.~C.-C. Chu, ``Digital asset
  management with distributed permission over blockchain and attribute-based
  access control,'' in \emph{IEEE International Conference on Services
  Computing (SCC)}, 2018, pp. 193--200.

\bibitem{jpdcref16}
N.~A. Popova and N.~G. Butakova, ``Research of a possibility of using
  blockchain technology without tokens to protect banking transactions,'' in
  \emph{IEEE Conference of Russian Young Researchers in Electrical and
  Electronic Engineering (EIConRus)}, 2019, pp. 1764--1768.

\bibitem{jpdcref17}
G.~M. Arantes, J.~N. D'Almeida, M.~T. Onodera, S.~M. D. B.~M. Moreno, and V.~D.
  R.~S. Almeida, ``Improving the process of lending, monitoring and evaluating
  through blockchain technologies: An application of blockchain in the
  brazilian development bank (bndes),'' in \emph{IEEE International Conference
  on Internet of Things (iThings)}, 2018, pp. 1181--1188.

\bibitem{challenge01}
D.~Desfontaines and B.~Pej{\'o}, ``Sok: Differential privacies,'' \emph{arXiv
  preprint arXiv:1906.01337}, 2019.

\end{thebibliography}

\vspace{0cm}
\begin{IEEEbiography}[{\includegraphics[width=1in,height=1.35in,clip,keepaspectratio]{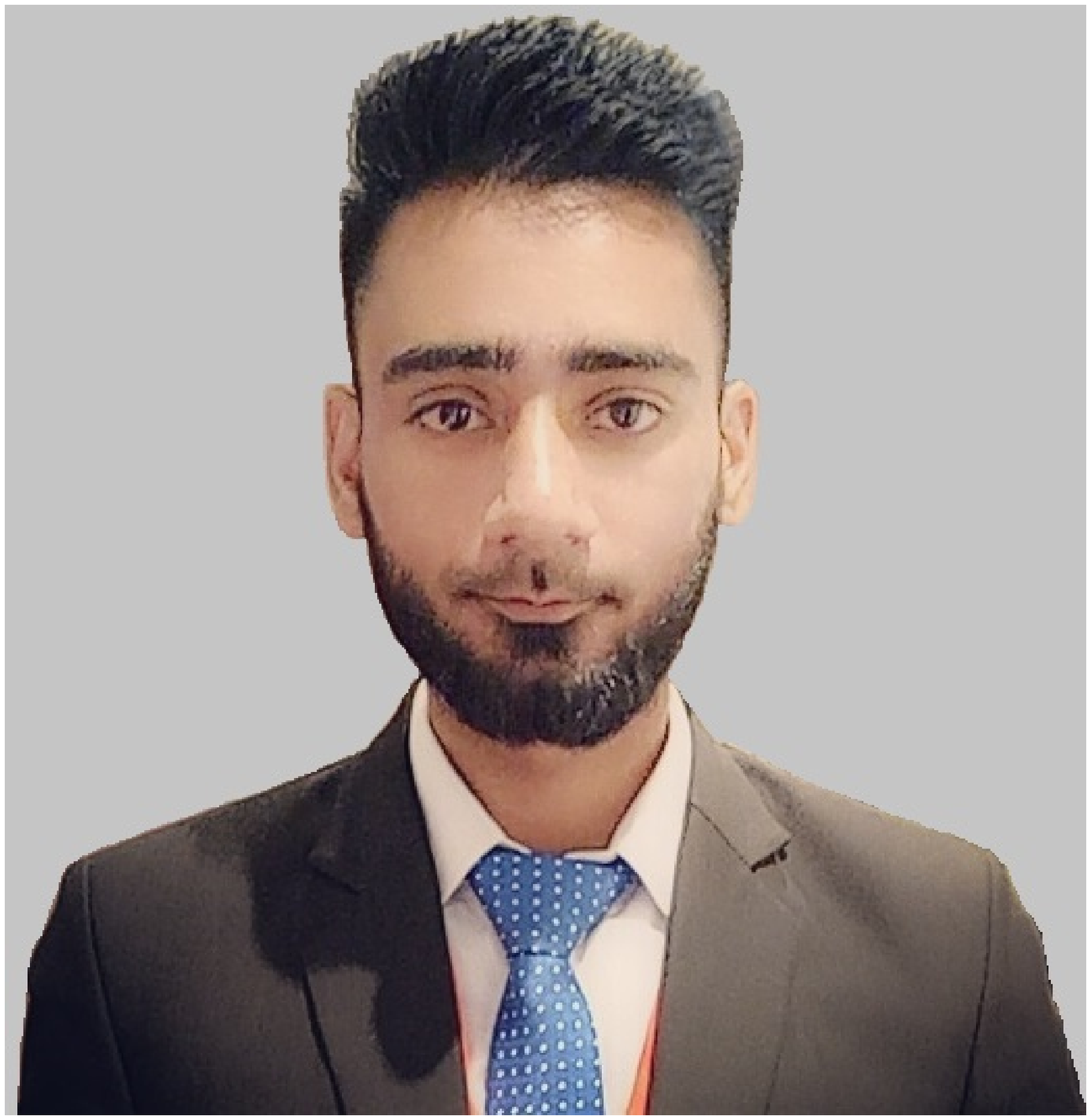}}]{Muneeb Ul Hassan}

received his Bachelor degree in Electrical Engineering from COMSATS Institute of Information Technology, Wah Cantt, Pakistan, in 2017. He received Gold Medal in Bachelor degree for being topper of Electrical Engineering Department. Currently, he is pursuing the Ph.D. degree from Swinburne University of Technology, Hawthorn VIC 3122, Australia. His research interests include privacy preservation, blockchain, game theory, and smart grid. He served in the TPC for the IEEE International Conference on Cloud Computing Technology and Science (CloudCom 2019). He is a reviewer of various journals, such as the IEEE Communications Surveys \& Tutorials, IEEE Journal on Selected Areas in Communications, Elsevier Future Generation Computing Systems, Journal of Network and Computer Applications, Computers \& Electrical Engineering, IEEE ACCESS, Wiley Transactions on Emerging Telecommunications Technologies, IEEE Journal of Communications and Networks, Springer Wireless Networks, Human-centric Computing and Information Sciences, and KSII Transactions on Internet and Information Systems. He also has been a Reviewer for various conferences, such as IEEE Vehicular Technology Conference (VTC)-Spring 2019, Vehicular Technology Conference (VTC)-Fall 2018, IEEE International Conference on Communications (ICC) - 2019, International workshop on e-Health Pervasive Wireless Applications and Services e-HPWAS’18, IEEE Globecom 2018 workshop: Security in Health Informatics (SHInfo2018), Frontiers of Information Technology 2019, Frontiers of Information Technology 2018. 

\end{IEEEbiography}
\vspace{0cm}

\begin{IEEEbiography}[{\includegraphics[width=1in,height=1.35in,clip,keepaspectratio]{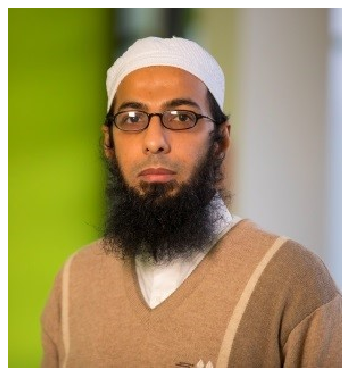}}]{Mubashir Husain Rehmani (M’14-SM’15)}

received the B.Eng. degree in computer systems engineering from Mehran University of Engineering and Technology, Jamshoro, Pakistan, in 2004, the M.S. degree from the University of Paris XI, Paris, France, in 2008, and the Ph.D. degree from the University Pierre and Marie Curie, Paris, in 2011. He is currently working as Assistant Lecturer at Cork Institute of Technology (CIT), Ireland. He worked at Telecommunications Software and Systems Group (TSSG), Waterford Institute of Technology (WIT), Waterford, Ireland as Post-Doctoral researcher from Sep 2017 to Oct 2018. He served for five years as an Assistant Professor at COMSATS Institute of Information Technology, Wah Cantt., Pakistan.  He is currently an Area Editor of the IEEE Communications Surveys and Tutorials. He served for three years (from 2015 to 2017) as an Associate Editor of the IEEE Communications Surveys and Tutorials. Currently, he serves as Associate Editor of  IEEE Communications Magazine, Elsevier Journal of Network and Computer Applications (JNCA), and the Journal of Communications and Networks (JCN). He is also serving as a Guest Editor of Elsevier Ad Hoc Networks journal, Elsevier Future Generation Computer Systems journal, the IEEE Transactions on Industrial Informatics, and Elsevier Pervasive and Mobile Computing journal. He has authored/ edited two books published by IGI Global, USA, one book published by CRC Press, USA, and one book with Wiley, U.K. He received “Best Researcher of the Year 2015 of COMSATS Wah” award in 2015. He received the certificate of appreciation, “Exemplary Editor of the IEEE Communications Surveys and Tutorials for the year 2015” from the IEEE Communications Society. He received Best Paper Award from IEEE ComSoc Technical Committee on Communications Systems Integration and Modeling (CSIM), in IEEE ICC 2017. He consecutively received research productivity award in 2016-17 and also ranked \# 1 in all Engineering disciplines from Pakistan Council for Science and Technology (PCST), Government of Pakistan. He also received Best Paper Award in 2017 from Higher Education Commission (HEC), Government of Pakistan.

\end{IEEEbiography}
\vspace{0cm}

\begin{IEEEbiography}[{\includegraphics[width=1in,height=1.25in,clip,keepaspectratio]{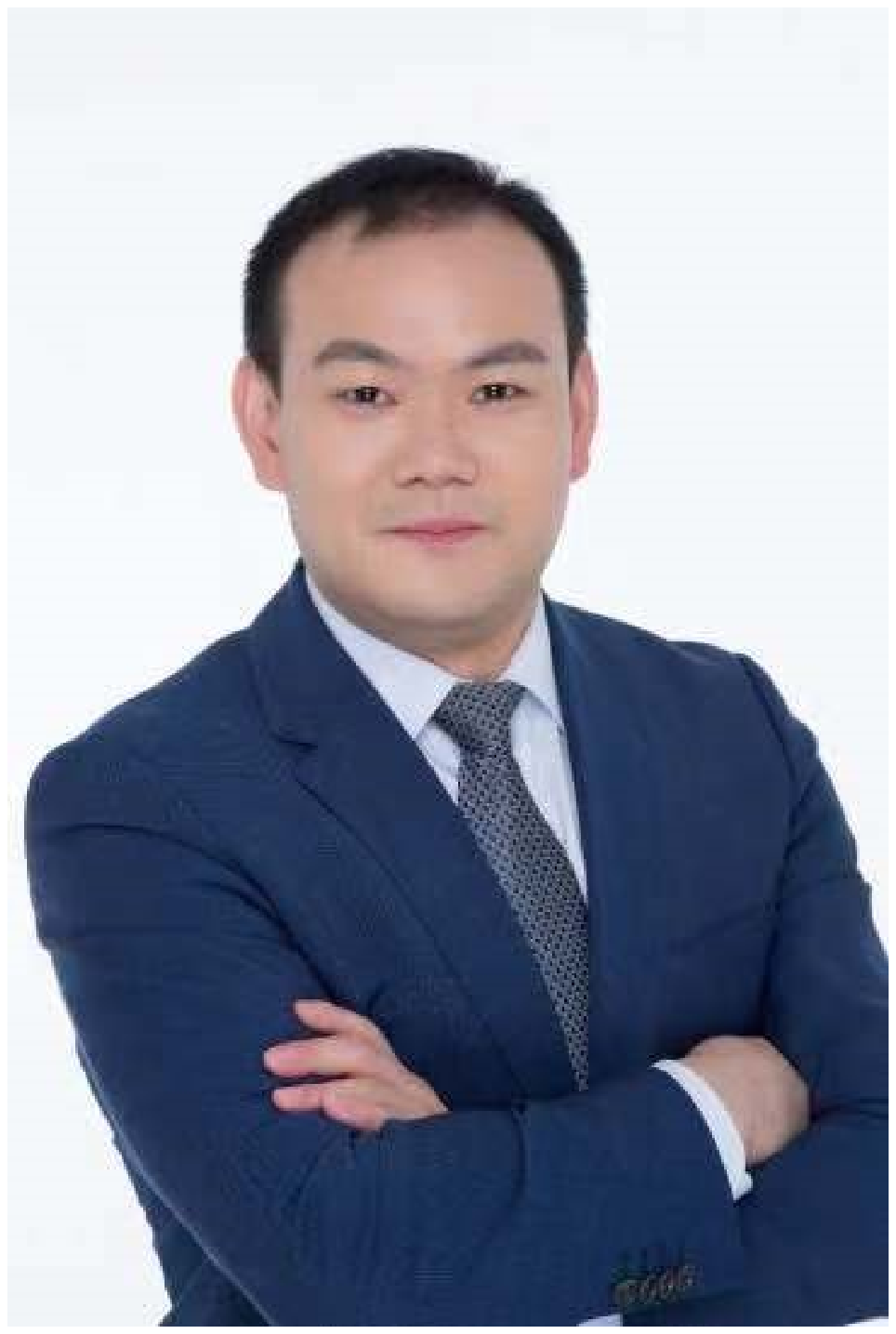}}]{Dr. Jinjun Chen}

is a Professor from Swinburne University of Technology, Australia. He is Deputy Director of Swinburne Data Science Research Institute. He holds a PhD in Information Technology from Swinburne University of Technology, Australia. His research interests include scalability, big data, data science, data systems, cloud computing, data privacy and security, health data analytics and related various research topics. His research results have been published in more than 160 papers in international journals and conferences, including various IEEE/ACM Transactions.  

\end{IEEEbiography}

\end{document}